\documentclass[10pt,journal,compsoc]{IEEEtran}

\usepackage[ruled]{algorithm2e}
\usepackage{amsmath}
\usepackage{amssymb}
\usepackage{blindtext}
\usepackage{graphicx}
\usepackage[mathlines,switch]{lineno}
\usepackage[numbers]{natbib}
\usepackage[tight,footnotesize]{subfigure}
\usepackage{units}

\usepackage{xcolor}
\newcommand\change[1]{#1}

\newenvironment{glist}[3]{
\begin{list}{#1}{\topsep 0.0mm
                        \partopsep 0.0mm
                        \setlength{\itemindent}{#3}
                        \parsep 0.0mm
                        \itemsep #2\baselineskip
                        \settowidth{\labelwidth}{#1}
                        \leftmargin 0.0mm
                        \addtolength{\leftmargin}{\itemindent}
                        \addtolength{\leftmargin}{\labelwidth}
                        \addtolength{\leftmargin}{\labelsep}}}{\end{list}}

\begin{document}


\title{GROUPS-NET: Group Meetings Aware Routing in Multi-Hop D2D Networks}

\author{
  Ivan~O.~Nunes,
  Clayson~Celes,
  Pedro~O.~S.~Vaz~de~Melo
  and Antonio~A.~F.~Loureiro
\IEEEcompsocitemizethanks{\IEEEcompsocthanksitem Ivan~O.~Nunes, Clayson~Celes, Pedro~O.~S.~Vaz~de~Melo and Antonio~A.~F.~Loureiro are with the Department of Computer Science from the Federal University of Minas Gerais, Belo Horizonte, MG, 31270-901.\protect\\
E-mail:  \{ivanolive, claysonceles, olmo, loureiro\}@dcc.ufmg.br
}
}

\date{5-8 July 2016}




\IEEEtitleabstractindextext{%
\begin{abstract}
Device-to-device (D2D) communication will allow direct transmission between nearby mobile devices in the next generation cellular networks. A fundamental problem in multi-hop D2D networks is the design of forwarding algorithms that achieve, at the same time, high delivery ratio and low network overhead. In this work, we study group meetings' properties by looking at their structure and regularity with the goal of applying such knowledge in the design of a forwarding algorithm for D2D multi-hop networks. We introduce a forwarding protocol, namely GROUPS-NET, which is aware of social group meetings and their evolution over time. Our algorithm is parameter-calibration free and does not require any knowledge about the social network structure of the system. In particular, different from the state of the art algorithms, GROUPS-NET does not need communities' detection, which is a complex and expensive task. We validate our algorithm by using different publicly available data sources. In real-world large scale scenarios, our algorithm achieves approximately the same delivery ratio of the state-of-the-art solution with 40\% less network overhead.
\end{abstract}

}

\maketitle




\section{Introduction}

In recent years, high data rate applications such as videos, songs, games, and social media have become increasingly popular in cellular networks. Device-to-Device (D2D) communication has been proposed to facilitate high data rate transmissions among near users offering higher throughput, efficient spectral usage, extended network coverage and improved energy efficiency.

\change{D2D refers to the direct transmission of content between devices in close proximity, without need for all data to go through the base station. To enable direct D2D \textit{ad hoc} links in the current cellular network architecture, the 3GPP consortium proposed Device-to-Device Proximity Service (D2D ProSe)~\cite{D2DProSe}. D2D ProSe reuses LTE radio interface for direct transmission between devices, allowing the base station to participate and control D2D connection establishment and creation of routes between devices. Several studies in the context of 3GPP-LTE~\cite{asadi2014survey} have shown D2D's great potential for alleviating data traffic demands. Nowadays, D2D spans way further, being considered not only as a 3GPP technology, but as one of the key enabling technologies in 5G cellular networks~\cite{5GRoadmap}.}

\begin{figure}[!t]
\centering
  \includegraphics[width=1.5in]{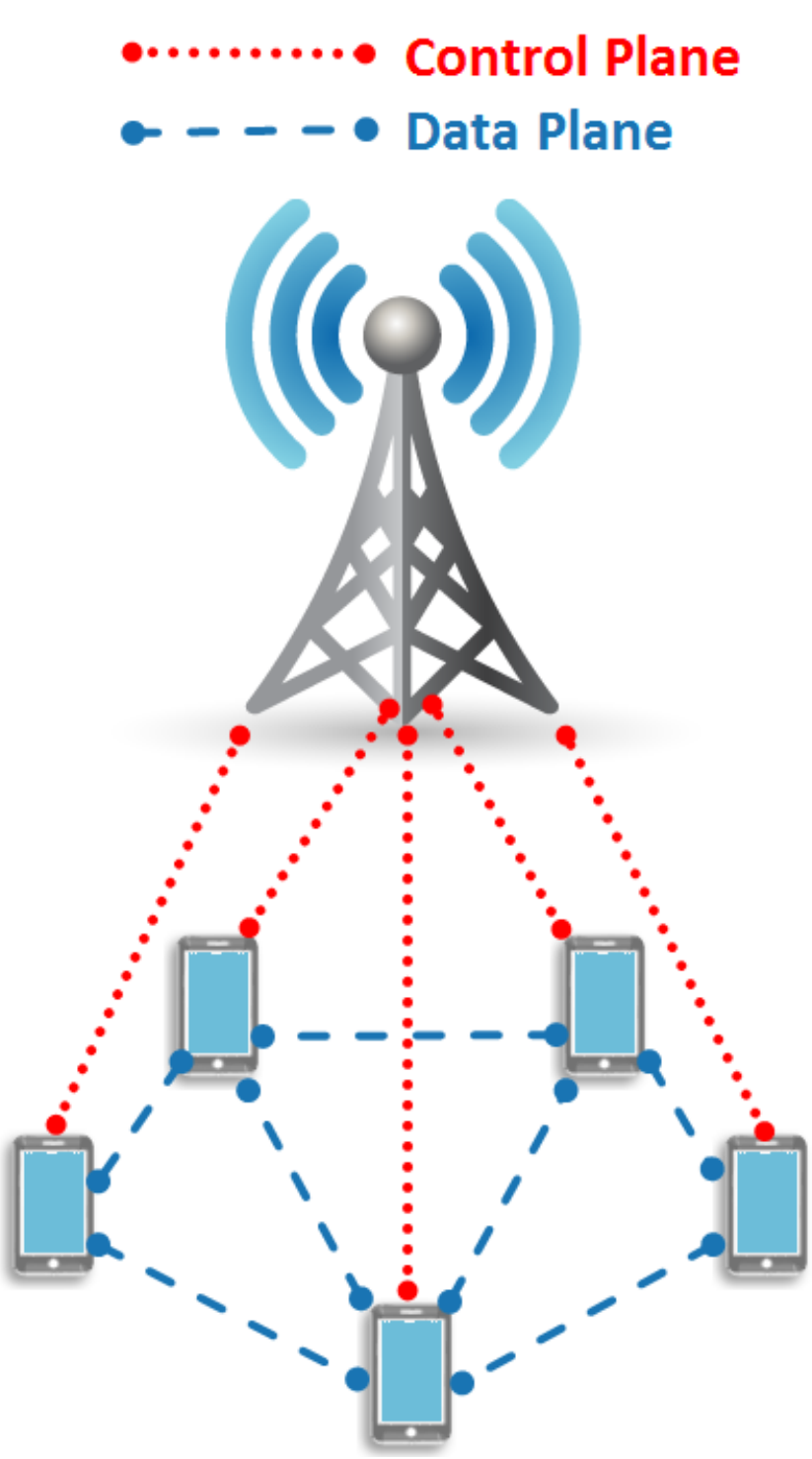}
\caption{D2D hybrid architecture: centralized control and distributed data planes}\label{fig1}
\end{figure}

D2D modes can be classified into two basic types: \textbf{1-hop transmission}, in which a message goes directly from the source to the destination if they are close enough to each other; and \textbf{multi-hop transmission}, where the message must be opportunistically routed, device by device, from the source to the destination. This last strategy is more complex, since it depends on the intermittent communication structure of a mobile network and is suited for communications that tolerate larger delivery times. This concept was initially introduced in the context of Delay Tolerant Networks (DTNs), but it has many applications to D2D networks as well~\cite{D2Dbubble}. A fundamental difference from D2D networks to pure DTNs is the possibility of a centralized control plane with a distributed data plan, as depicted in Figure~\ref{fig1}. In this case, the base station controls the forwarding policy while the data is transmitted from device-to-device until it reaches the destination~\cite{wcm}.

Forwarding algorithms in multi-hop D2D networks aim to achieve cost-effective delivery, i.e., the highest possible delivery ratio with the lowest possible network overhead. In this case, the delivery ratio is measured as the percentage of messages routed opportunistically that are successfully delivered to the destination. Successfully delivered messages are the ones that the base station will not need to deliver itself, enabling bandwidth offload. The network overhead is measured by the average number of times that the content needs to be D2D-transmitted for the message to reach its destination. A high number of transmissions may negatively impact the users' experience by, for example, increasing the devices' energy expenditure.

Considering these metrics, the most successful strategy for opportunistic cost-effective forwarding in D2D Networks, Bubble Rap~\cite{bubble,D2Dbubble}, relies on two aspects: information about the nodes' centrality and the static social communities they are part of. The first element can be approximated by the node popularity within the mobile network. The latter, however, presents some significant problems. First, communities  are computationally expensive to identify~\cite{mobicom2011}. Second, they are hard to detect in a distributed way, since individual nodes do not have information about the contact graph of the whole network. Existing distributed community detection algorithms achieve at most 85\% precision in the detected communities~\cite{dist_com}. Another problem of community detection algorithms is the parameter calibration, which depends on the parameters to be adjusted for each given scenario~\cite{parameters}. In a real-time application, such as D2D communication, such calibration is not feasible. Moreover, there is no established truth for community detection. In fact, Abrahao et {al.}~\cite{separability} showed that, for the same scenarios, different community detection algorithms yield very different results for communities' compositions. Finally, static communities detection does not account for the dynamism in humans' social relationships, i.e., how they change over time.

Given all these issues, in this work we go beyond and propose the use of social groups' meetings instead of communities. A social group meeting is defined as \textit{a group of people who are together, in space and time, for some social reason or common goal}. For example, friends hanging out at a bar share the social motivation of being together to relax and talk to each other. Students in the classroom share the objective of learning the class' subject content. People in a bus are together because they share the same goal of going to the same route. These are examples of social group meetings. As human beings have regular schedules and routines, it is reasonable to expect social group meetings to present some regularity as well.

From the implementation point of view, different from communities, detecting a social group meeting in a distributed fashion is an easy task. A device can detect a group meeting of which it is part of by simply looking at the list of devices that remained nearby for more than a given time, for example, \unit[10]{min}. Moreover, the group meeting detection method does not change depending on the scenario nor require parameters' calibration for each specific network, as in community detection schemes. In addition to those desirable characteristics, by looking only at recent group meetings or by giving a higher importance to more recent meetings, it is possible to account for the dynamic nature of social relationships. All of these favorable characteristics motivated us to employ group meetings in the design of an opportunistic routing scheme that is better suited for D2D networks than the current social-aware proposals. \change{Generally speaking, \textit{\textbf{the main contribution of the present work is to introduce social group meetings and their regularity as a better metric for social context, specially for opportunistic D2D communication and \textit{ad hoc} networks, in which community detection is a hard task}}. Specifically, the contributions of this work are fourfold:}
\begin{glist}{$\bullet$}{0}{0mm}
\topsep 0mm
\parskip 0mm
\partopsep 0mm
\parindent 0mm
\itemsep 0mm
\parsep 0mm
\item A methodology for detecting and tracking group meetings from pairwise contact traces;
\item A characterization and a model of group meetings, which can be used to predict future meetings using the information about the most recent ones;
\item An analysis of the state-of-the-art synthetic mobility models, which allowed us to conclude that group meetings' properties are not well captured by the synthetic traces generated from such models;
\item An opportunistic forwarding algorithm for D2D networks, aware of group meetings, which does not need to detect communities nor calibrate parameters, and that achieves better cost-effectiveness than the state-of-the-art solution in real large-scale scenarios.
\end{glist}

This paper is organized as follows. Section~\ref{sec:RW} presents the related work and discusses the corresponding contributions with respect to human mobility,  cost-effective opportunistic routing, and D2D communication. Section~\ref{sec:GMD} formalizes our methodology to detect and track group meetings from pairwise contact traces. Section~\ref{properties} describes the main properties of group meetings that make them interesting in the design of a new forwarding strategy. Section~\ref{groups} introduces GROUPS-NET, a group meetings aware routing protocol. Section~\ref{sec:SRM} presents a comparison of real and synthetic mobility traces regarding the presence of group mobility. Section~\ref{sec:results} evaluates GROUPS-NET, contrasting its performance with the state-of-the-art solution, Bubble Rap, in different network scales. Finally, Section~\ref{sec:conslusion} presents final remarks and future work.

\section{Related Work}\label{sec:RW}

\textbf{Human mobility.} Many studies have used diverse data sources to look into human mobility patterns in the perspective of both individual behavior and collective dynamics. Gonzalez et al.~\cite{individual_mob} show that human trajectories present a high degree of temporal and spatial regularity. More recently, a dichotomy in individual mobility was revealed in~\cite{PappalardoNature2015} and~\cite{SongNature2010}. Those studies suggest that two mobility profiles, called returners and explorers, govern people movements based on preferential returns and explorations of new places. In relation to collective dynamics, Candia et al.~\cite{candia2008uncovering} analyze large-scale collective behavior from aggregated call detail records. Their work reveal that the spatio-temporal fluctuations of individuals in a city strongly depend on the activity patterns and routines. Yet, Issacman et al.~\cite{IsaacmanMobiSys2012} propose an approach for modeling how people move in different metropolitan areas.

All the studies above are concerned with identifying the intrinsic properties of human mobility in order to provide knowledge and models for different applications, such as the impact of large-scale events in urban mobility~\cite{events}, typical transitions between points of interest in a city~\cite{2014revealing}, and characterization and prediction of traffic conditions~\cite{traffic2,traffic1,traffic3}. Some studies have investigated the importance of understanding the properties of human mobility for designing communication protocols based on opportunistic encounters among people~\cite{ChaintreauTMC2007, PanissonAdHocNetworks2012, RaoTMC2015, SermpezisTMC2015}. These studies explore pairwise contacts between users from the individual mobility perspective considering the following metrics: contact rate, inter-contact time, and contact duration. However, to the best of our knowledge, there is no such characterization related to group mobility and its applicability to mobile networks, which is one of the contributions of this work.

\textbf{Opportunistic Routing.} Among several solutions proposed for opportunistic routing, the most successful are the probabilistic and social-aware routing. Probabilistic routing in DTNs was introduced by Lindgren et al.~\cite{prophet}. The central idea of their approach is that pairwise encounters that happened more recently are more likely to happen again in the near future. The PRoPHET algorithm, which implements this idea, achieved great success, being years later outperformed by social-aware strategies. Hui et al.~\cite{bubble}, for instance, use the social community structure, detected from contact graphs of mobile networks, combined with network nodes' centrality, to propose a forwarding algorithm named Bubble Rap. Although there are other routing protocols that exploit the social information from contacts between people~\cite{DalyMobihoc2007, HossmannINFOCOM2010, MtibaaINFOCOM2010}, to the best of our knowledge, until now Bubble Rap has been the most cost-effective forwarding algorithm in terms of high delivery ratio and low network overhead. A version of  Bubble Rap for D2D networks was proposed in~\cite{D2Dbubble}. As mentioned earlier, community detection has several problems that undermine its application to real networks. In the present work, we introduce GROUPS-NET, a social-aware forwarding algorithm based on the tracking of social group meetings, which does not suffer from such problems and achieve better cost-effectiveness than Bubble Rap in real-world large scale scenarios.

\change{\textbf{D2D Communication.} In this work we explore group meetings to enable opportunistic routing in multihop D2D networks. Previous efforts explore D2D in several ways to provide different applications. Goratti et al.~\cite{goratti2014connectivity} propose a D2D protocol for public safety use cases, such as natural disasters with massive people concentration. The protocol includes device discovery through beacon broadcast using encryption keys to set up secure D2D communication links. Some works have shown the potential of using LTE-D2D to provide communication in infrastruture-less settings, such as in wearable networks~\cite{steri2016lte} and in the internet of things~\cite{steri2016novel}. Vehicular networks may also benefit from the adoption of D2D. In~\cite{li2016traffic}, the authors propose a traffic flow model based on a dynamic resource allocation algorithm to deal with the problem of vehicular communication, sharing the same licensed frequency spectrum as D2D communications, in a single LTE cell. In~\cite{lee2016performance}, the combination of LTE-D2D with MIMO is proposed to achieve low latency and better reliability in vehicular communication.}

\section{Detecting and Tracking Group Meetings}\label{sec:GMD}

Group meetings may be easily detected in a distributed manner in a real scenario by, for instance, looking at the list of near devices. However, to study group meetings' properties it is necessary to detect group meetings from pairwise contacts traces, which are the typical available data sources to study social-aware forwarding algorithms. In such traces, each pairwise contact records the two nodes involved and the time when the contact happened. In this section, we describe the methodology we used to detect group meetings from pairwise contact traces. Notice that in a real distributed scenario these steps would not be necessary since group meetings detection is simple to perform in a distributed manner. However, this methodology must be applied to enable such study using pairwise contact traces.

In the present study, we used the MIT Reality Mining~\cite{mit} and Dartmouth~\cite{dartmouth} traces, which are contact traces containing 80 and 1200 users, respectively. In the MIT Reality Mining,  the monitored users reside in two university buildings and were monitored for several months. Contacts were registered when two users were less than 10 meters apart. Although the MIT Reality Mining trace consists of a specific and small scale scenario, we considered this trace in the present work since it is the original trace used to validate Bubble Rap in~\cite{bubble}. The Dartmouth trace registered contacts of all students in a university campus for two months. To the best of our knowledge, Dartmouth is the largest scale and publicly available contact dataset. Due to its scale and generality, the Dartmouth trace is a better representation of a real D2D cellular network environment.

\subsection{Modeling Contact Traces as Graphs}\label{model}

To detect group meetings, we propose a social graph model. First, we slice the proximity dataset in time windows $tw$. We discuss the ideal time size for $tw$ in Section~\ref{data_cha}. Contacts within the same slice $tw$ will be aggregated in a contact graph $Gc(V,E[tw = i])$, in which $V$ is the set of vertices representing entities in the data set (i.e., people) and $E$ is the set of edges that represent proximity contacts between a pair of entities in $V$. Thus, in our model, the trace processing will result in a set of subsequent, undirected, edge-weighted graphs: $S = \{Gc(V,E[tw=0]),Gc(V,E[tw=1]), \ldots, Gc(V,E[tw =n])\}$. The weight of an edge $vw$ $\in$ $E$ is given by (i) the number of contacts registered between $v$ and $w$ during the time slice $tw$; or (ii) the sum of $vw$ contacts duration, inside $tw$ time slice, if contacts' duration is available in the trace. Since in the MIT Reality Mining this information is not available, we used the first option. For the Dartmouth trace, we used the second option because contacts' duration is available.

\subsection{Data Characterization}\label{data_cha}

\begin{figure*}[!t]
\centering
  \subfigure[]{\includegraphics[width=2.2in]{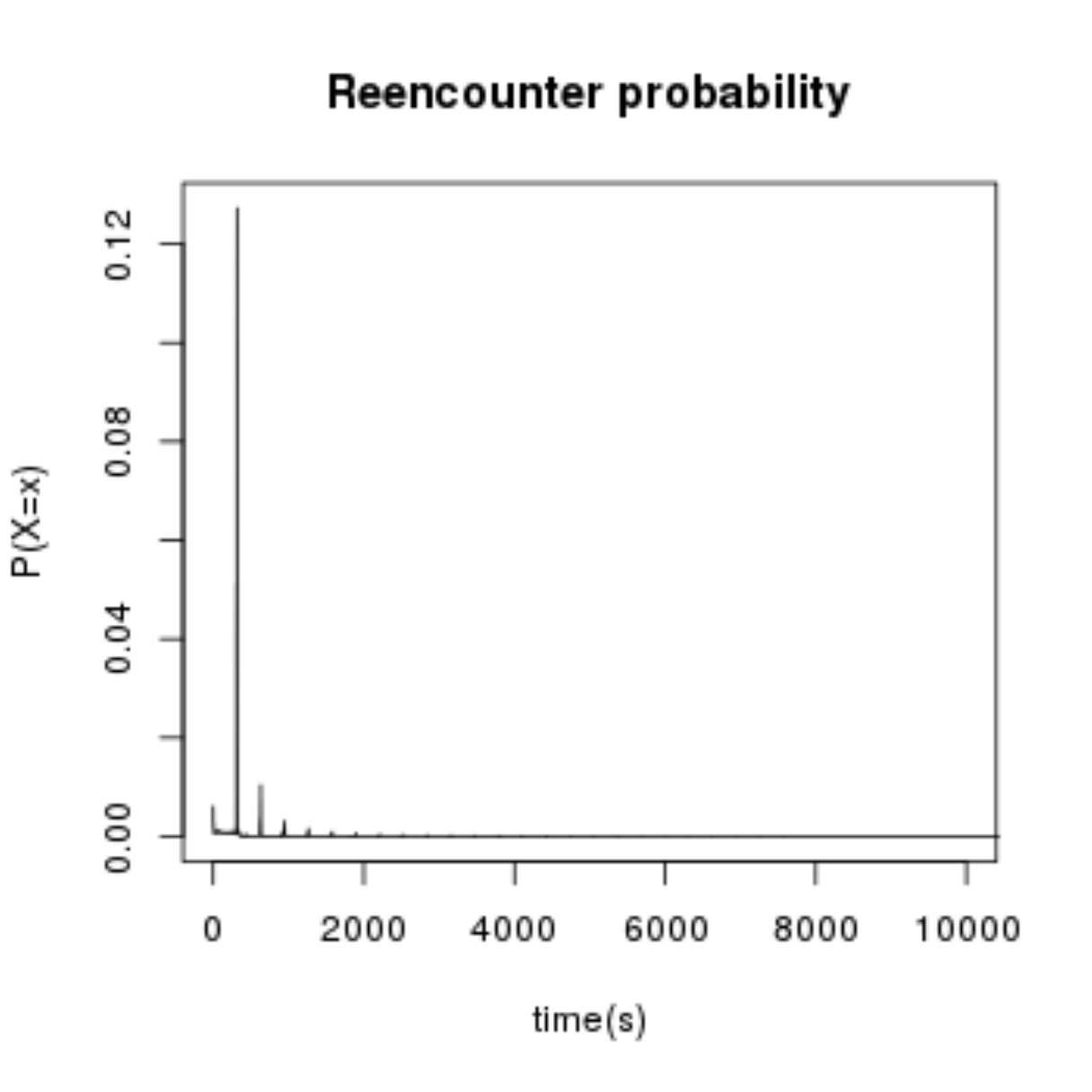}\label{fig_first_case}}
  \hfil
  \subfigure[]{\includegraphics[width=2.2in]{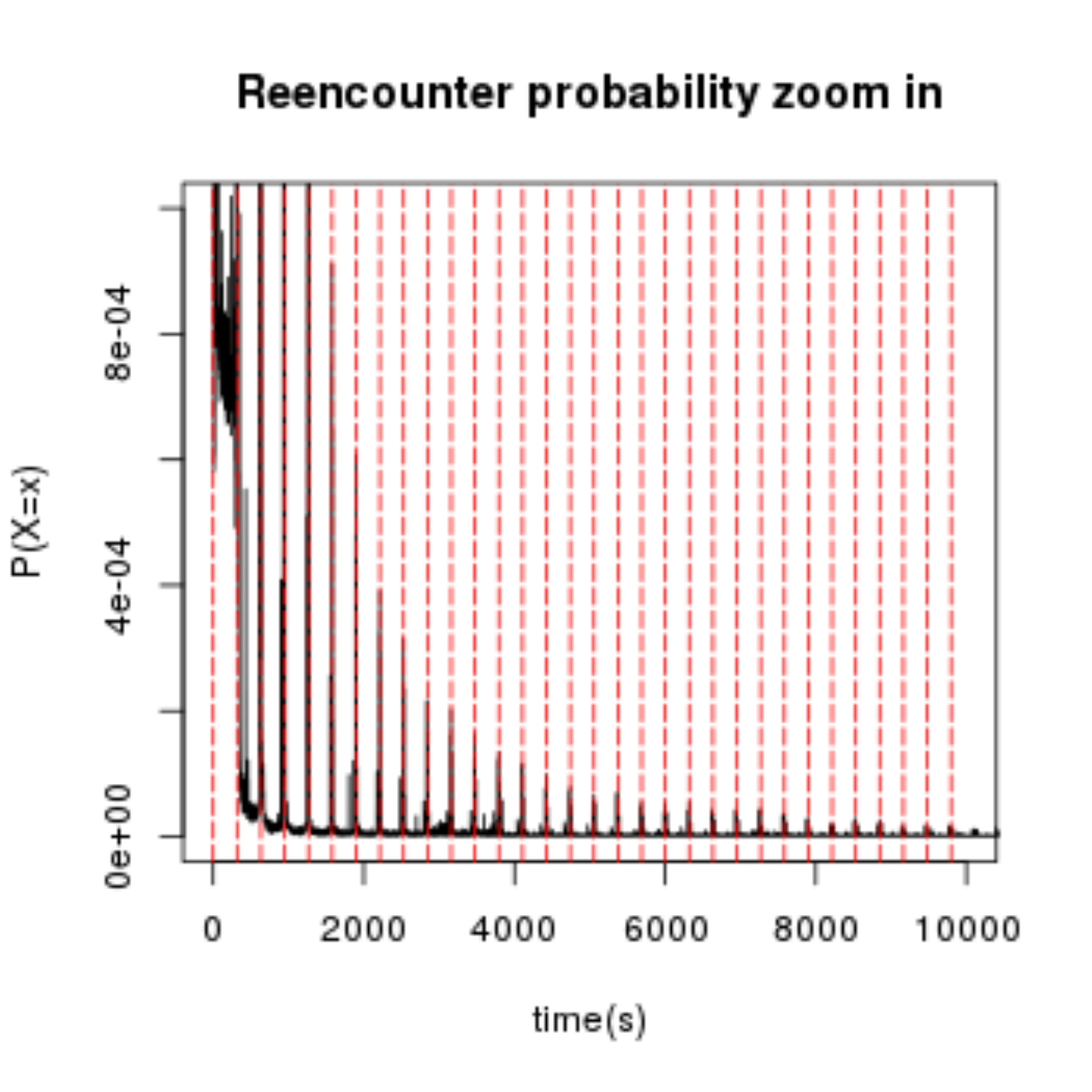}\label{fig_second_case}}
  \hfil
  \subfigure[]{\includegraphics[width=2.2in]{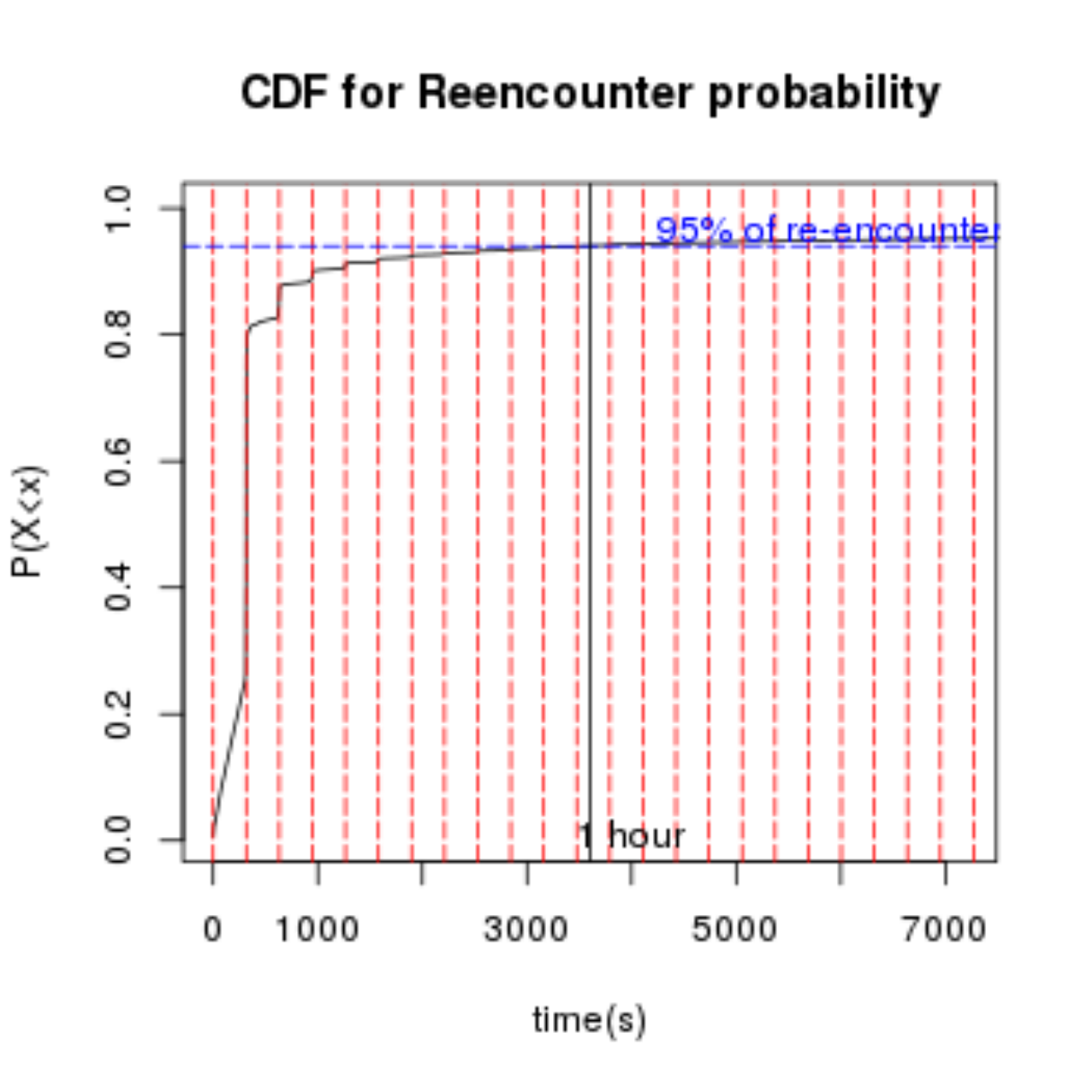}\label{fig_third_case}}
\caption{Probability function of the time $x$ (in seconds) until the next meeting. Red dashed lines show a fixed inter-measurement time of 318 seconds in which re-encounter peaks happen. This means that most of the trace proximity records were acquired in fixed periods of 318 seconds. 95\% of the pairwise re-encounters happen in less than one hour}\label{fig_reenc}
\end{figure*}

Considering a pre-defined $tw$ size, e.g., \unit[30]{min}, a single contact within the entire time slice does not necessarily mean that entities are socially interacting. Single contacts might be random encounters, even if nodes share a social bond. It might be caused by intersections between individual trajectories and should not necessarily mean a social interaction. We are interested in defining a threshold $w_{th}$ for the minimum edge weight, which should be enough to consider that the entities are in fact together inside the time slice $tw$. It is clear that to properly define both the size of slice $tw$ and the edge weight threshold $w_{th}$ we need to analyze the data set properties (e.g., sampling rate). In this section, we perform a characterization of the studied datasets to be able to properly define $tw$ and $w_{th}$. Due to space limitation, we only present the characterization of the MIT Reality Mining trace, but the same methodology can be applied to the Dartmouth trace and any other contact dataset as well.

\begin{figure}[!t]
\centering
  \includegraphics[width=\columnwidth]{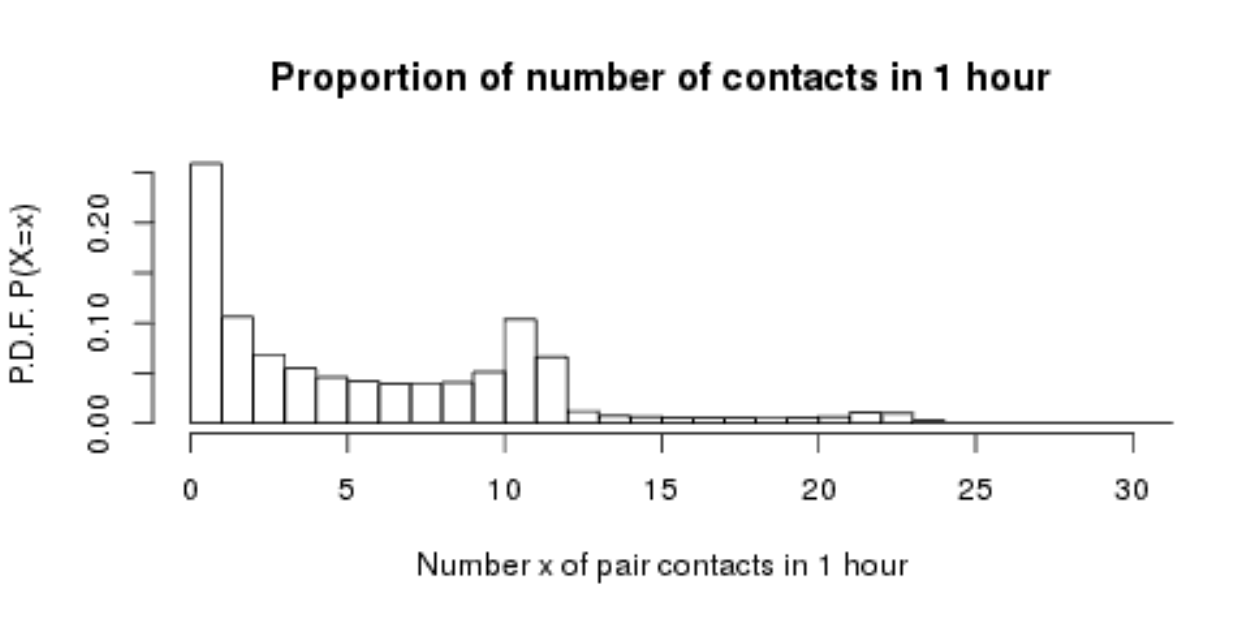}
\caption{Probability distribution function of the number of pair contacts per hour}\label{hour_contacts}
\end{figure}

\begin{figure*}[!t]
\centering
  \subfigure[6AM]{\includegraphics[width=2.3in]{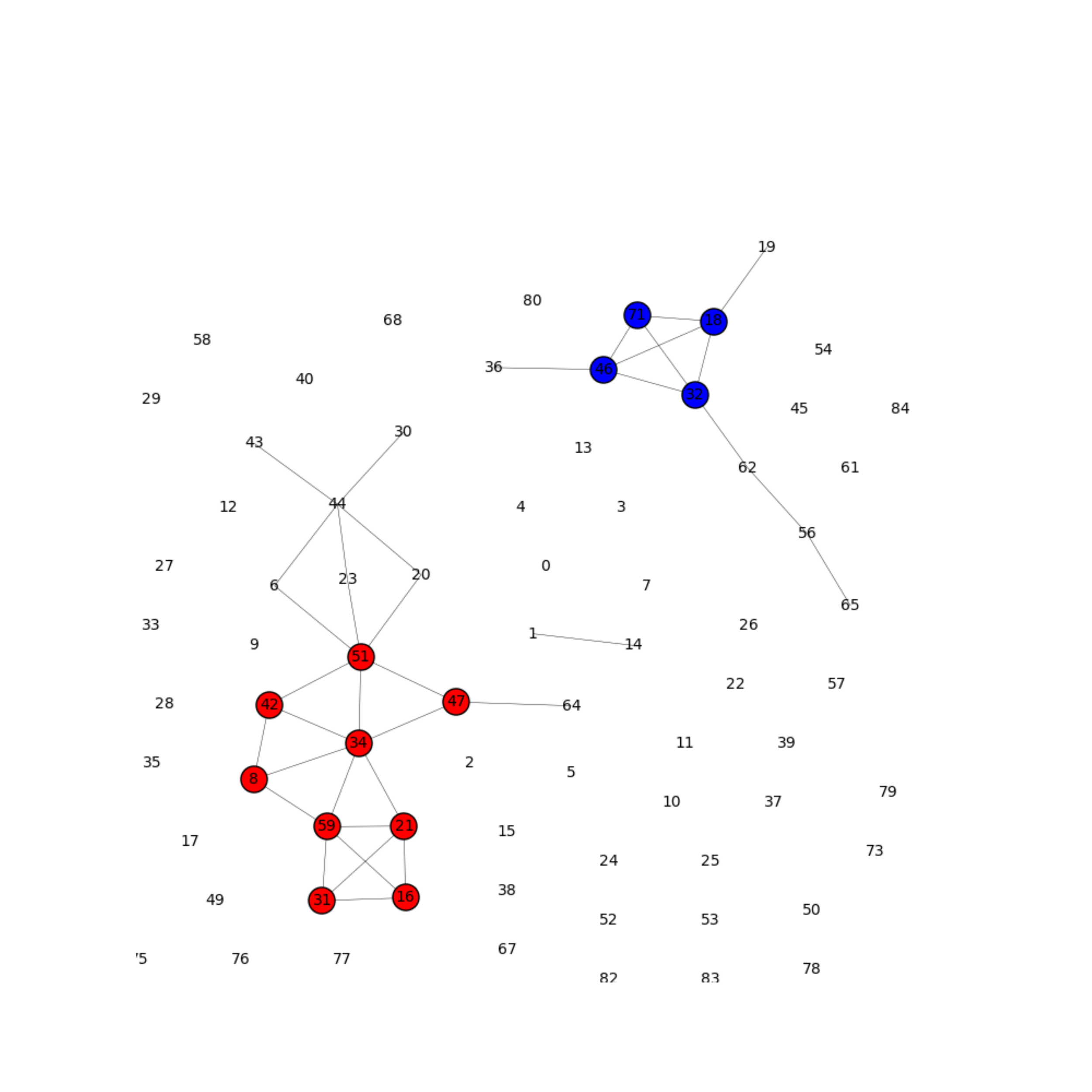}}
  \subfigure[7AM]{\includegraphics[width=2.3in]{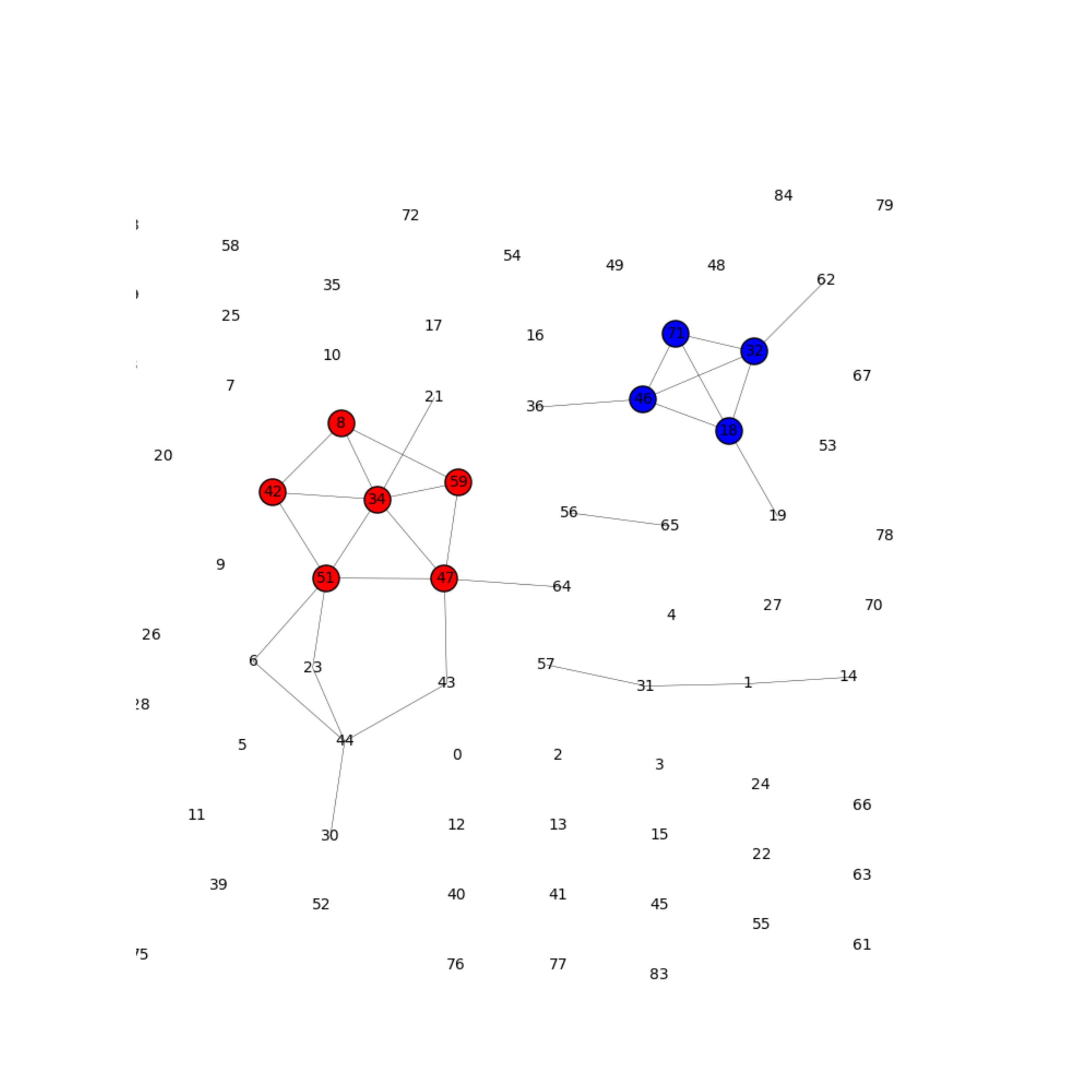}}
  \subfigure[8AM]{\includegraphics[width=2.3in]{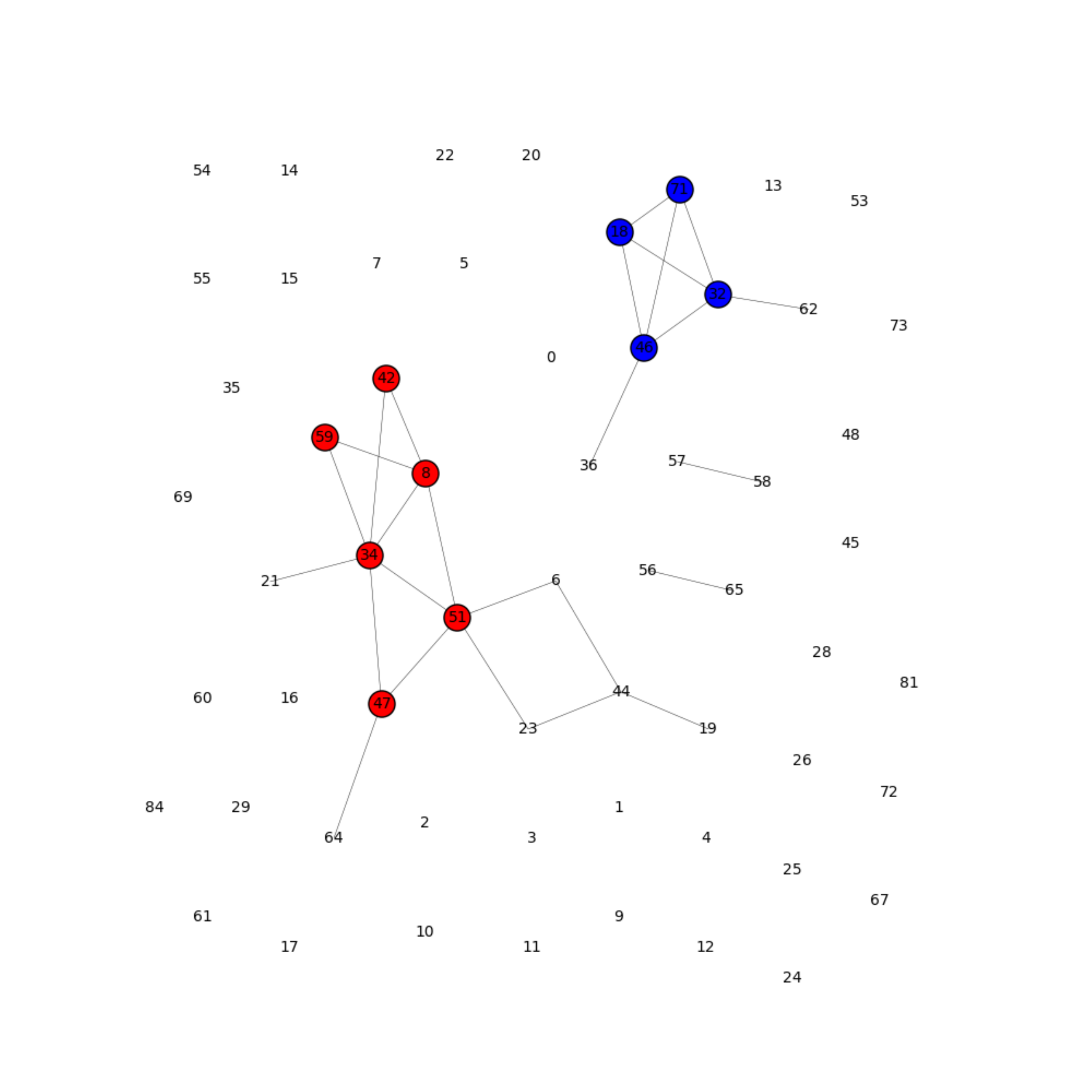}}
  \subfigure[9AM]{\includegraphics[width=2.3in]{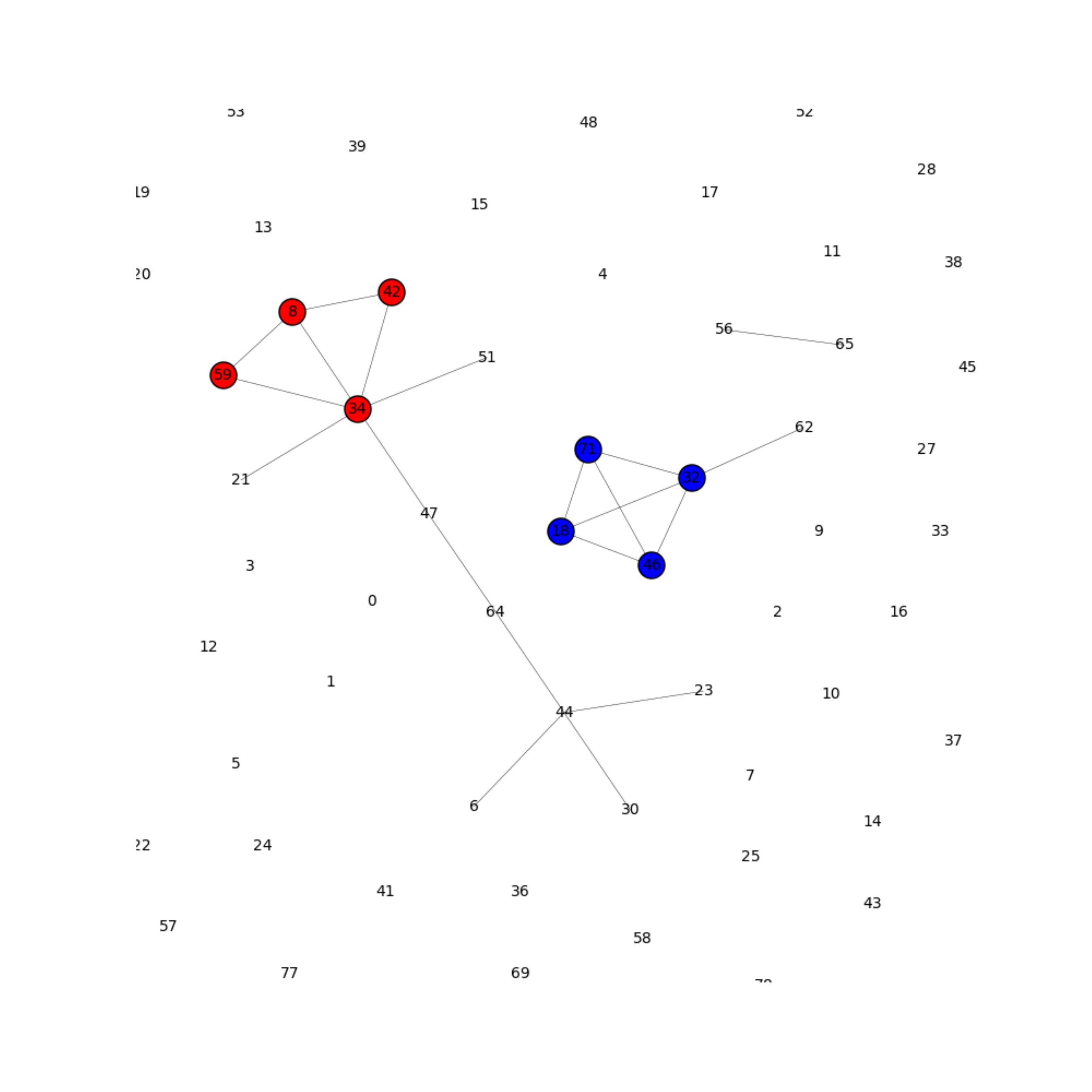}}
  \subfigure[10AM]{\includegraphics[width=2.3in]{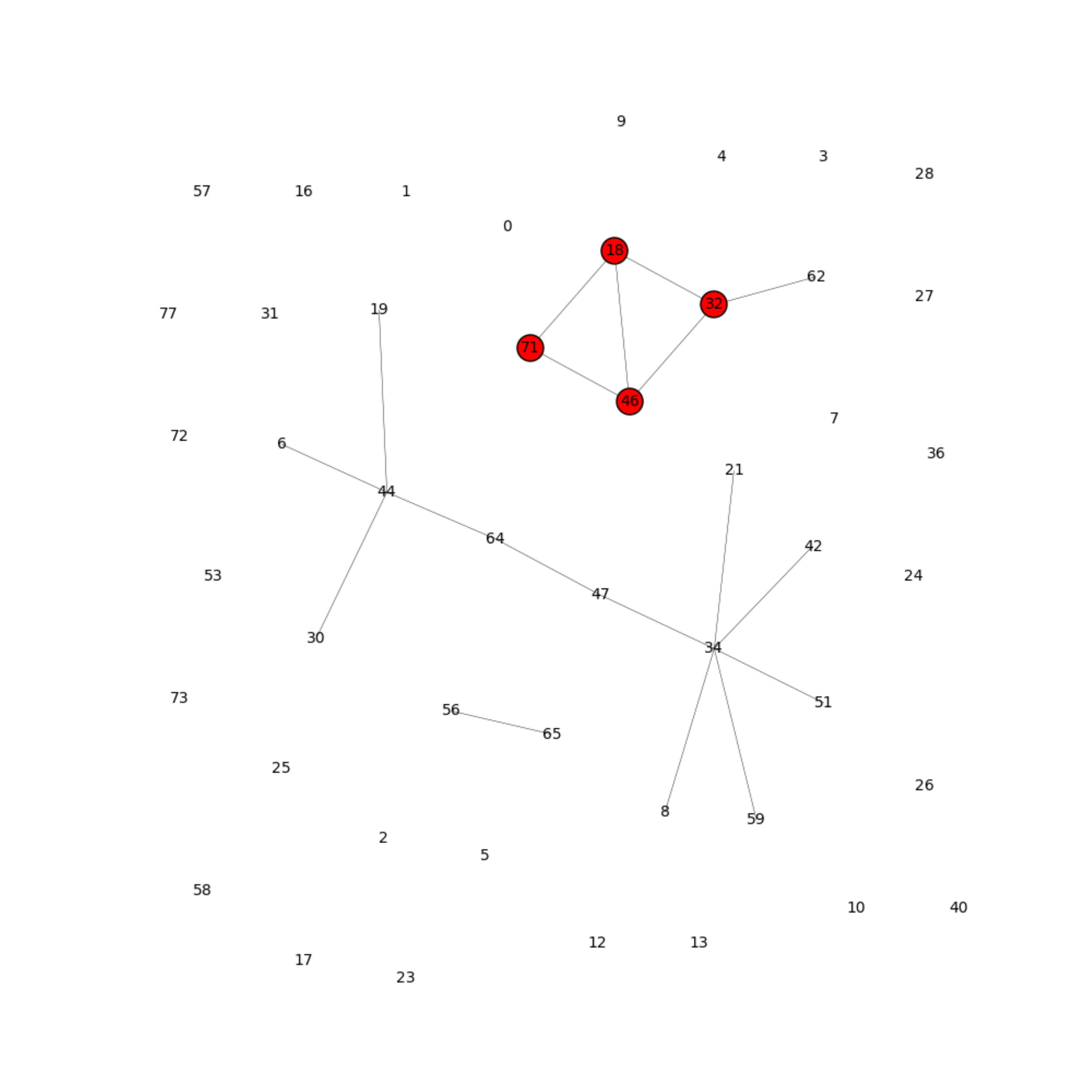}}
\caption{Group detection with CPM, in the MIT proximity trace, with $tw = $1h in three consecutive time windows, at February 5th of 2009. Only edges with $w_{th} \geq 2$ are represented}\label{mit_groups}
\end{figure*}

First, we analyze the time between pair re-encounters, i.e., once a pair has met, what is the distribution of the time until the next meeting. Figure~\ref{fig_reenc} shows that the re-encounter behavior is very periodical, with peaks around every \unit[5]{min} (red dashed lines). This behavior indicates that the deployed system for data acquisition acts at every \unit[5]{min} most of the times, but  it can also actuate in shorter periods. Looking at the CDF of the re-encounter probability, approximately 95\% of the re-encounters can be captured with a $tw$ of one hour. Thus, we set the duration of the time window $tw$ to one hour.

\change{Next, we analyze the frequency distribution of all contact pairs that happen within the same hour to define $w_{th}$. Figure~\ref{hour_contacts} shows that 27\% of pairs who meet in a given hour only meet once. We assume these one-time meetings as coincidence meetings, i.e., intersections in individual trajectories that do not imply social interaction. For meeting frequencies from 2 to 12, Figure~\ref{hour_contacts} shows more uniform values between 5\% and 10\%. For frequencies higher than 12, the probability becomes very low, which is consistent with the data acquisition rate that happens mostly in periods of \unit[5]{min} and rarely in less than five. From 2 to 12 encounters per hour, we have similar values in the CDF when compared to $P(X = 1)$. Based on this characterization, we define $w_{th} = 2$ for the MIT Reality Mining dataset, to filter non-social contacts from our group detection algorithm. In summary we are able to define that two or more contacts within an hour are enough to be considered social interaction in the MIT Reality Mining trace. Through similar analysis we have defined $tw =$ \unit[1]{h} and $w_{th} =$ \unit[10]{min} for the Dartmouth trace.}

\subsection{Detection and Tracking}\label{cpm}

After defining values for $tw$ and $w_{th}$, we define a social group as follows:
\begin{glist}{$\bullet$}{0}{0mm}
\topsep 0mm
\parskip 0mm
\partopsep 0mm
\parindent 0mm
\itemsep 0mm
\parsep 0mm
\item{\textbf{\textit{Definition of social group meeting:}}}\textit{ A group meeting is a community detected in $Gc(V,E[tw=i])$ , i.e., the graph generated from the $i^{th}$ time slice of the trace $S$, after eliminating edges between pairs with weight below the threshold $w_{th}$.}\footnote{As mentioned in the previous section, in practice groups' meetings would be distributively detected by devices, without the use of community detection.}
\end{glist}

So far, we have established a model to represent social interactions that consist of graphs generated from peer contacts in traces' time slices. Following the above group definition, we must be able to detect communities (social groups represented by more densely interconnected parts within a graph of social links) in such graphs in order to track social groups. There are several community detection algorithms, such as~\cite{copra,mobicom2011,c1}. In this work we, use the Clique Percolation Method (CPM)~\cite{palla2005} for two reasons: (i) community members can be reached through well connected subsets of nodes, and (ii) communities may overlap (share nodes with each other). This latter property is essential, as most social graphs are characterized by overlapping and nested communities~\cite{palla2007}. For each time-slice graph $Gc(V,E[tw=i])$, we compute CPM.

In CPM, a community is defined as a union of all $k$-cliques (complete sub-graphs of size $k$) that can be reached from each other through a series of adjacent $k$-cliques (where adjacency means sharing $k-1$ nodes). The CPM parameter $k$ limits the minimum size of detected communities. CPM has remarked itself as one of the most effective methods once fed with correct parameters~\cite{parameters}. We set $k=3$ to detect social groups, i.e., we consider groups of three or more people. Figure~\ref{mit_groups} shows groups detected with CPM in the MIT proximity trace at February 5th, 2009, from 6 to \unit[10]{am}, representing subsequent time slices of one-hour size.

Figure~\ref{mit_groups} shows that composition of groups may change over time, but some of the detected groups are present throughout several consecutive time windows $tw$. Since we are interested in investigating the regularity in group meetings, there must be a way of tracking them, i.e., a criterion for considering that two groups in consecutive time slices are in fact the same group. With that goal in mind, we introduce the Group Similarity Coefficient $\rho(G1,G2)$:
\begin{equation}\label{stability}
    \rho(G1,G2) = \frac{|V(G1) \cap V(G2)|}{|V(G1) \cup V(G2)|},
\end{equation}
where $|V(G1) \cap V(G2)|$ is the number of common nodes in groups $G1$ and $G2$ and $|V(G1) \cup V(G2)|$ is the total number of  different nodes present in the union set of both groups. The coefficient $\rho$ assumes values from 0 to 1, where 0 means no similarity, i.e., no node belongs to both groups, and 1 means that $G1$ and $G2$ have the same node composition. The group similarity coefficient is a measurement of the stability in groups' composition. We consider two time separated groups $G(tw=i)$ and $G(tw=j)$ to be the same if $\rho(G(tw=i),G(tw=j)) > 0.5$, i.e., if at least 50\% of the group members remain the same. Having a threshold of $\rho > 0.5$ simplifies the analysis, because otherwise groups would potentially divide in two groups with half of the original composition each, adding complexity to the analysis. At the same time, a $\rho$ threshold of $0.5$ allows high volatility in group composition, making it possible to better analyze the groups' evolution.

\subsection{\change{Group Meetings vs. Community Detection}}

 \change{The methodology we have presented throughout Sec. 3 is to detect social group meetings from contact traces. In other words, a way to deal with the dataset that we need to use in order to validate the idea we propose in this work. In a practical distributed scenario, specially within D2D communication domain, group detection is a much simpler task. In practice group meetings are easy to detect in a distributed fashion and communities are hard. On the other hand, group detection is hard to perform from contact traces, because user's location are not available in such traces and that is why we need the methodology we have described so far.}

 \change{One of the main drawbacks of using community detection in mobile networks, is that it requires offline parameter calibration. Moreover, community detection methods rely on global knowledge of the contact network, which is not practical in a large-scale distributed network. Group meetings detection, on the other hand, can rely solely on the discovery of nearby devices.}
 
 \change{For the reasons stated above, we here advocate group meetings awareness as a feasible and more realistic social-context metric for D2D communication than community detection. Distributed group detection can be achieved by simply using standard D2D peer-discovery network function to keep track of other devices that remain nearby for more than a period T. Even though T is a parameter, we argue that the value of T doesn't require fine offline tuning as k-clique sizes in community detection methods, for instance. The only requirement is that T is long enough so that one can really consider that  the nodes remained together and not just crossed each others path. In a practical scenario T would play the same role as $w_{th}$ in our methodology. Thus, the GROUPS-NET algorithm, that we discuss in section~\ref{properties}, is, to the best of our knowledge, the first social-aware forwarding scheme that does not require offline parameter calibration.}

\section{Social Group Meetings Properties}\label{properties}

In the previous section, we showed how  to detect group meetings from proximity traces. This section presents the main properties of  group meetings that could be used to leverage  D2D Routing. Further characterization of other group mobility properties may be found in~\cite{icc}.

\begin{figure*}[!t]
\centering
  \subfigure[]{\includegraphics[width=\columnwidth]{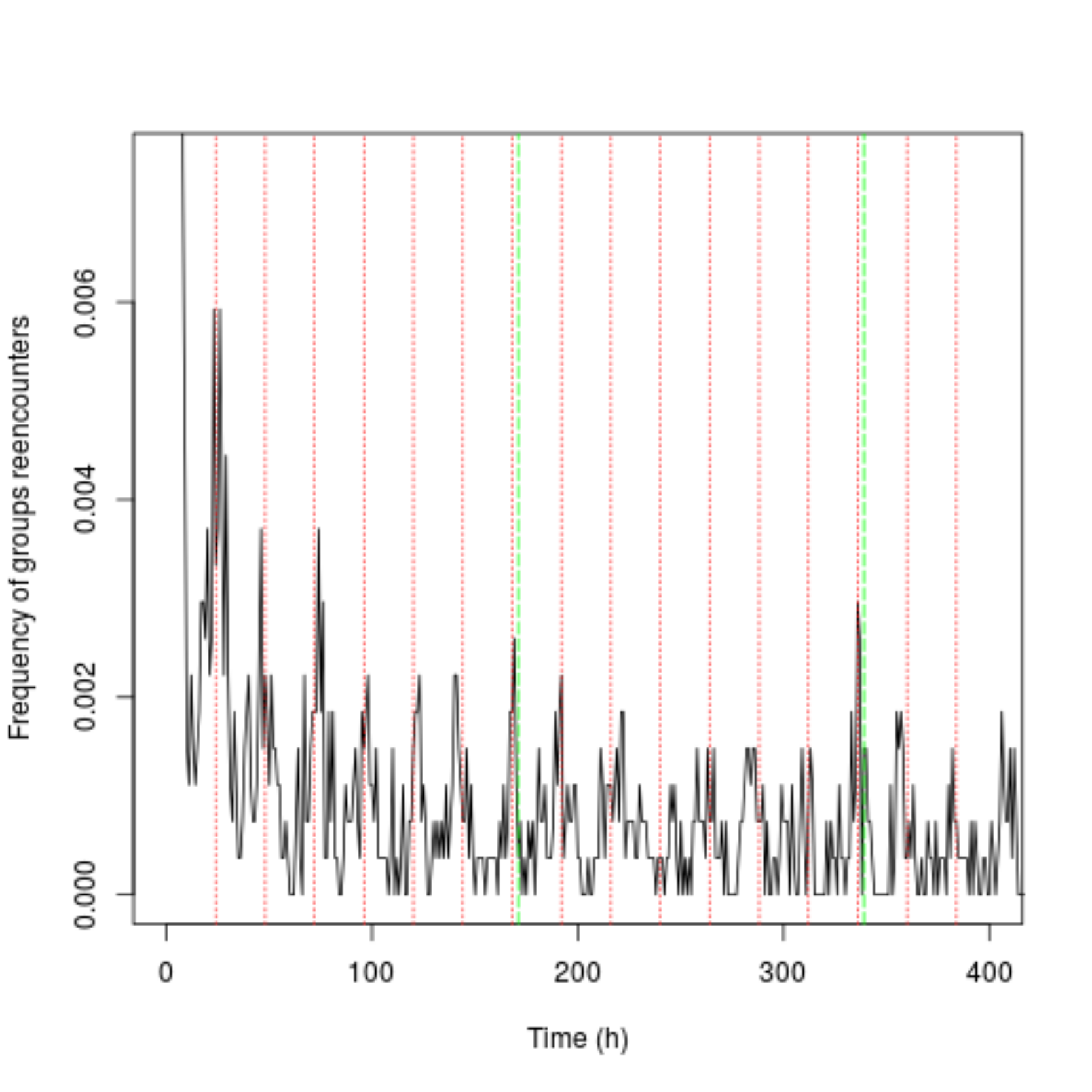}\label{b1}}
  \subfigure[]{\includegraphics[width=\columnwidth]{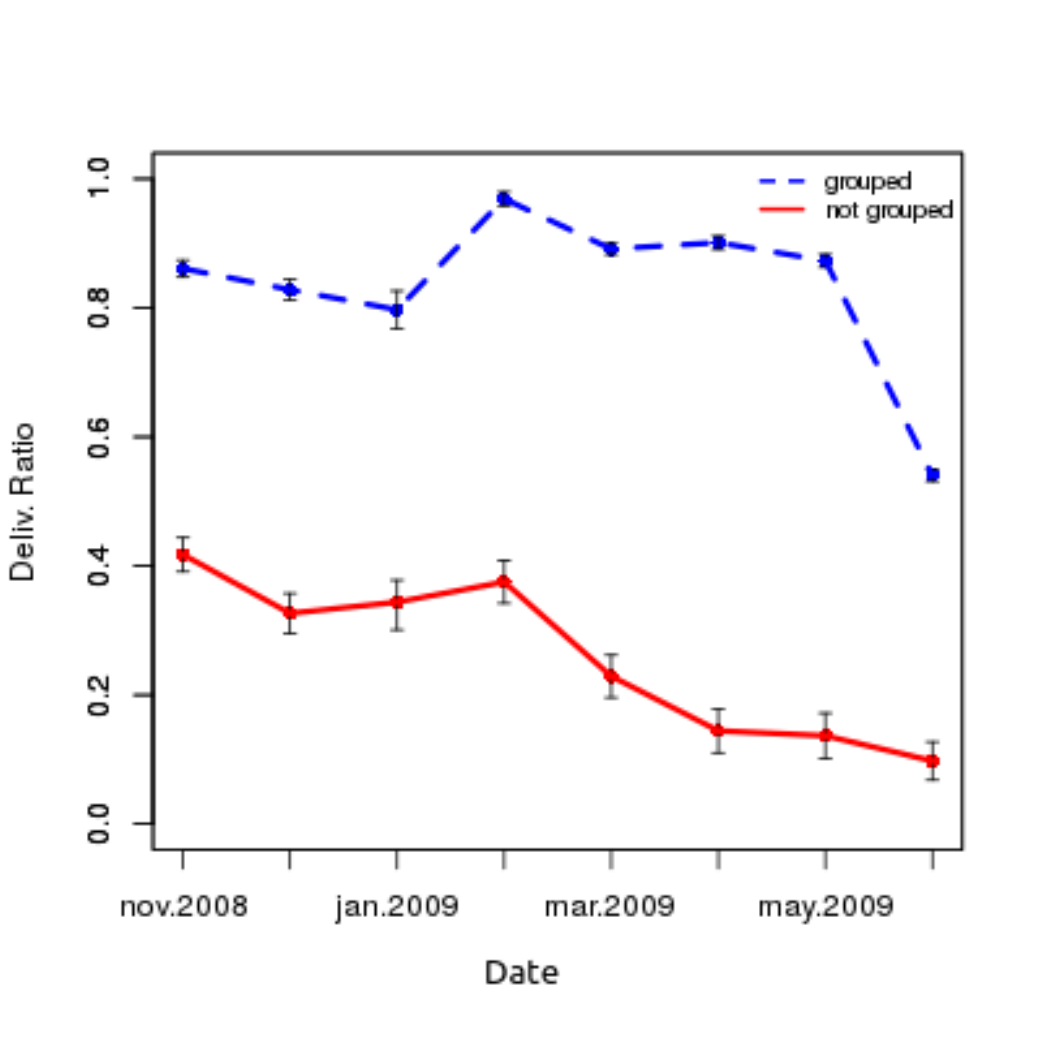}\label{b2}}
\caption{\textbf{a)} Probability of a given group re-meeting $t$ hours after its first meeting. Red dotted lines represent 24-hour periods and green dashed lines 7-day periods. \textbf{b)} Average delivery ratio, for different origin nodes, from November of 2008 to June of 2009}\label{motivation}
\end{figure*}

Figure~\ref{b1} presents the frequency of group re-encounters for the MIT Reality Mining, i.e., given the fact that a group first met at time $t=0$, how group re-meetings are distributed along the next hours ($t$ hours after the first meeting). The result reveals that the probability mass is concentrated around peaks of 24-hour periods (represented by red dotted lines). This means that group meetings are highly periodical in a daily fashion. One may also notice higher peaks marked with green dashed lines. Green dashed lines represent periods of seven days, meaning that groups' meetings also present weekly periodicity. This periodicity makes sense since people have schedules and routines. This result motivated our next experiment, which tries to answer the question: is it possible to use past group meetings to predict future ones?

In our next experiment, we select a node as the origin of a message and simulate an epidemic message transmission, i.e., every time a node with a message meets another node that does not have it yet, the message is propagated. We simulate the message propagation selecting each node of the MIT dataset as origin and divide the rest of the nodes into two classes: nodes that have belonged to a group together with the origin in the past 30 days and nodes that have not. Then, we compute the delivery ratios of the two classes of nodes. We consider that the message is delivered to node $N$ if node $N$ receives the message within seven days after the start of the dissemination.

As presented in Figure~\ref{b2}, for different months, the delivery ratios to nodes that have been in group meetings with the origin are over two times higher than of the other nodes. Around 90\% of the nodes that were in group meetings together with the origin received the message within one week. On the other hand, the delivery ratio to nodes that had not been in a group together with the origin is around 40\%. This result conforms with the periodical behavior presented in Figure~\ref{b1} (a group that has met in recent past is likely to meet again soon) and is a key insight on how group meetings could and should be used to better design opportunistic routing protocols. One may notice that in June the delivery ratios for both classes significantly drop. This behavior is explained by the fact that the trace was collected in a university campus and, in June, most students in the US start to leave the campus for summer vacation.

\begin{figure*}[!t]
\centering
  \subfigure[]{\includegraphics[width=\columnwidth]{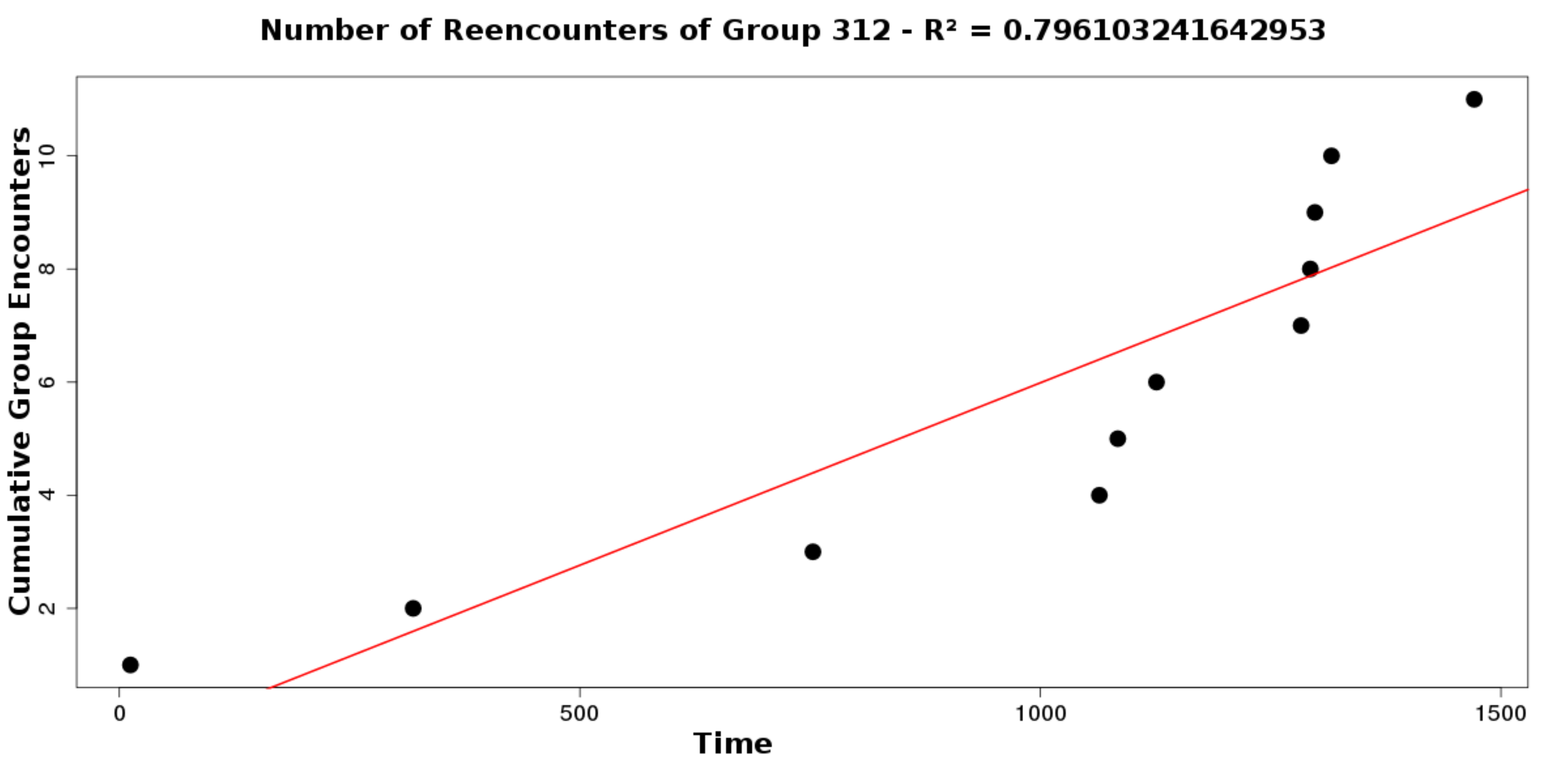}\label{p1}}
  \subfigure[]{\includegraphics[width=\columnwidth]{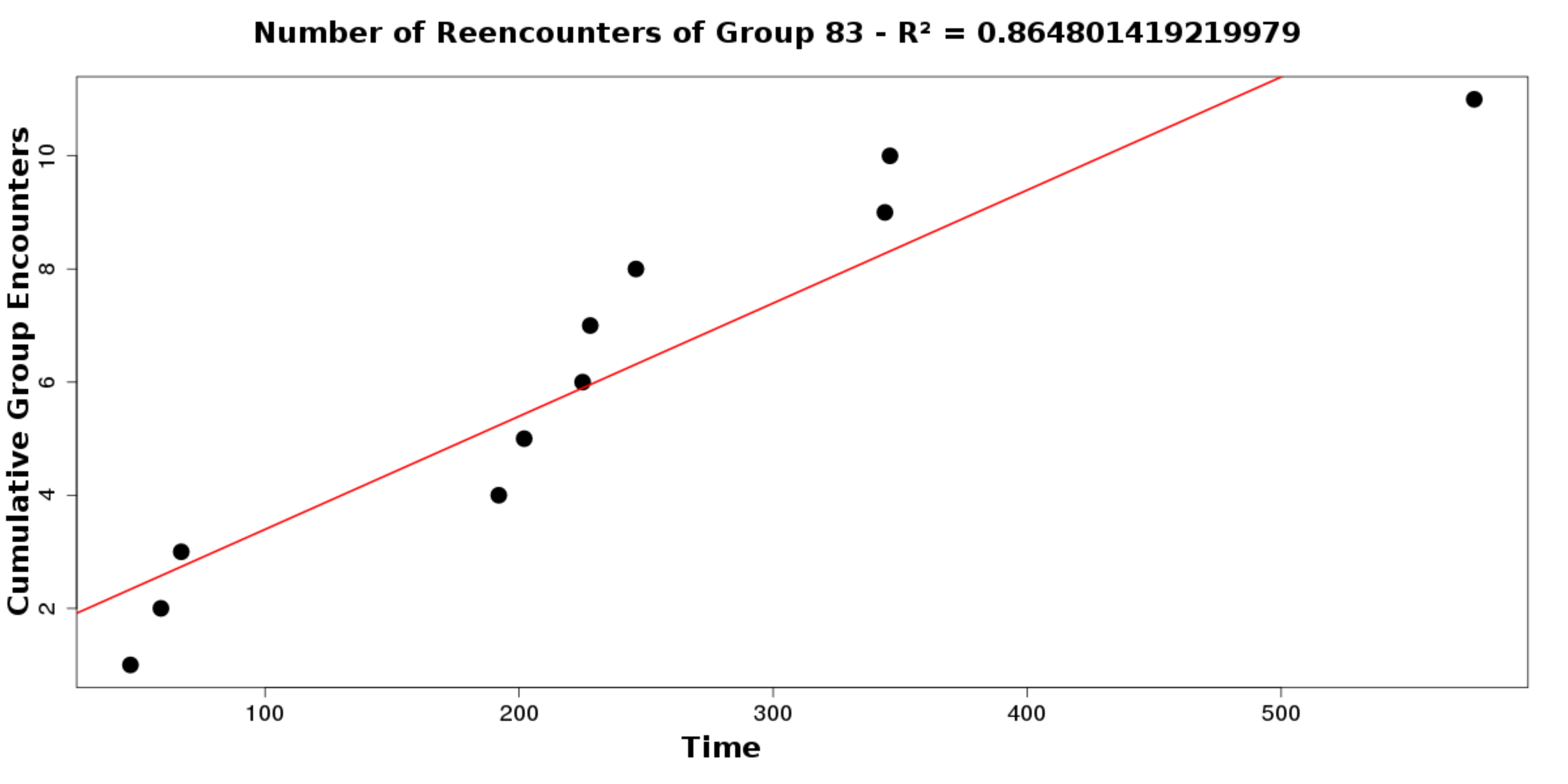}\label{p2}}
  \subfigure[]{\includegraphics[width=\columnwidth]{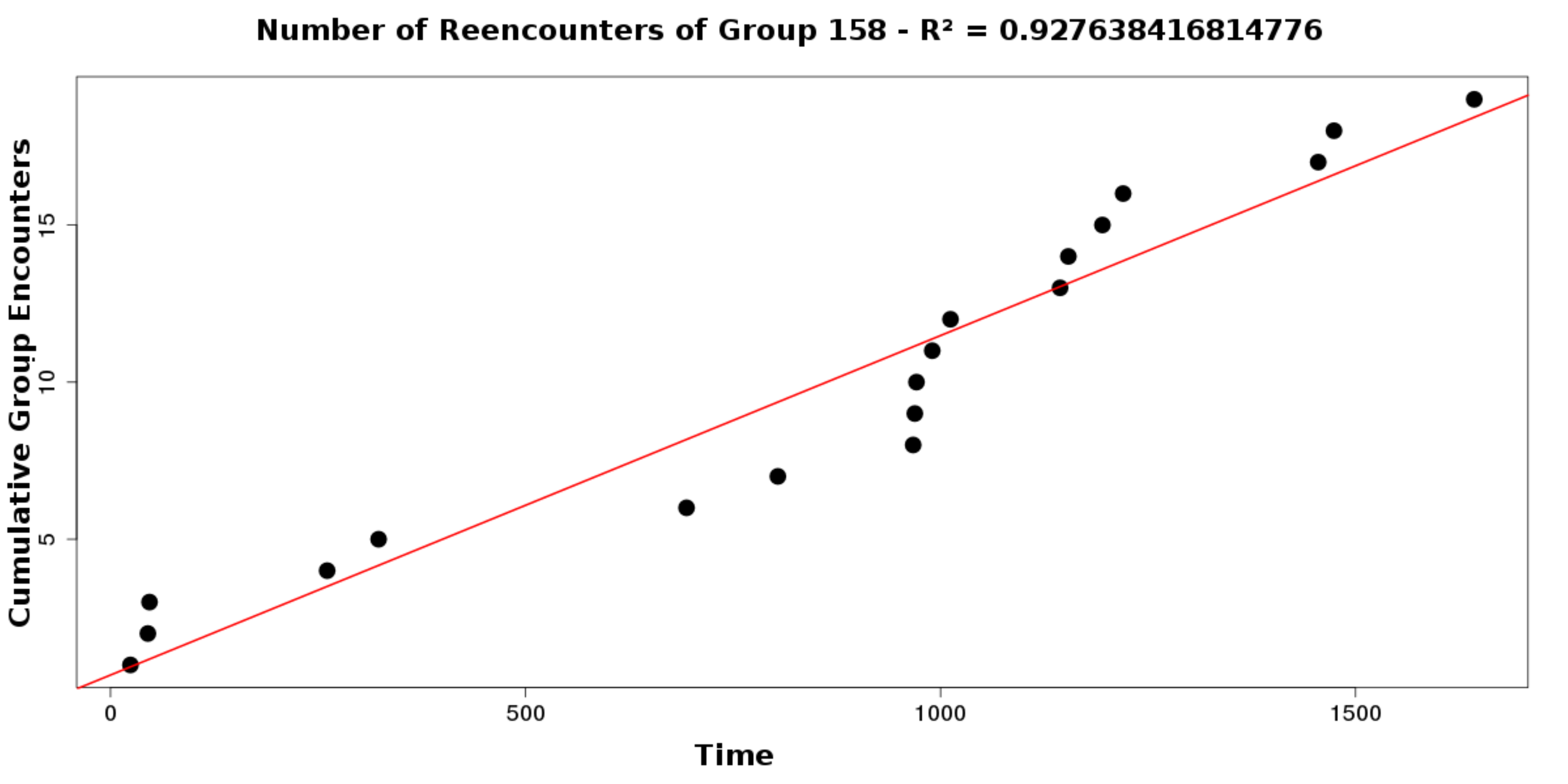}\label{p3}}
  \subfigure[]{\includegraphics[width=\columnwidth]{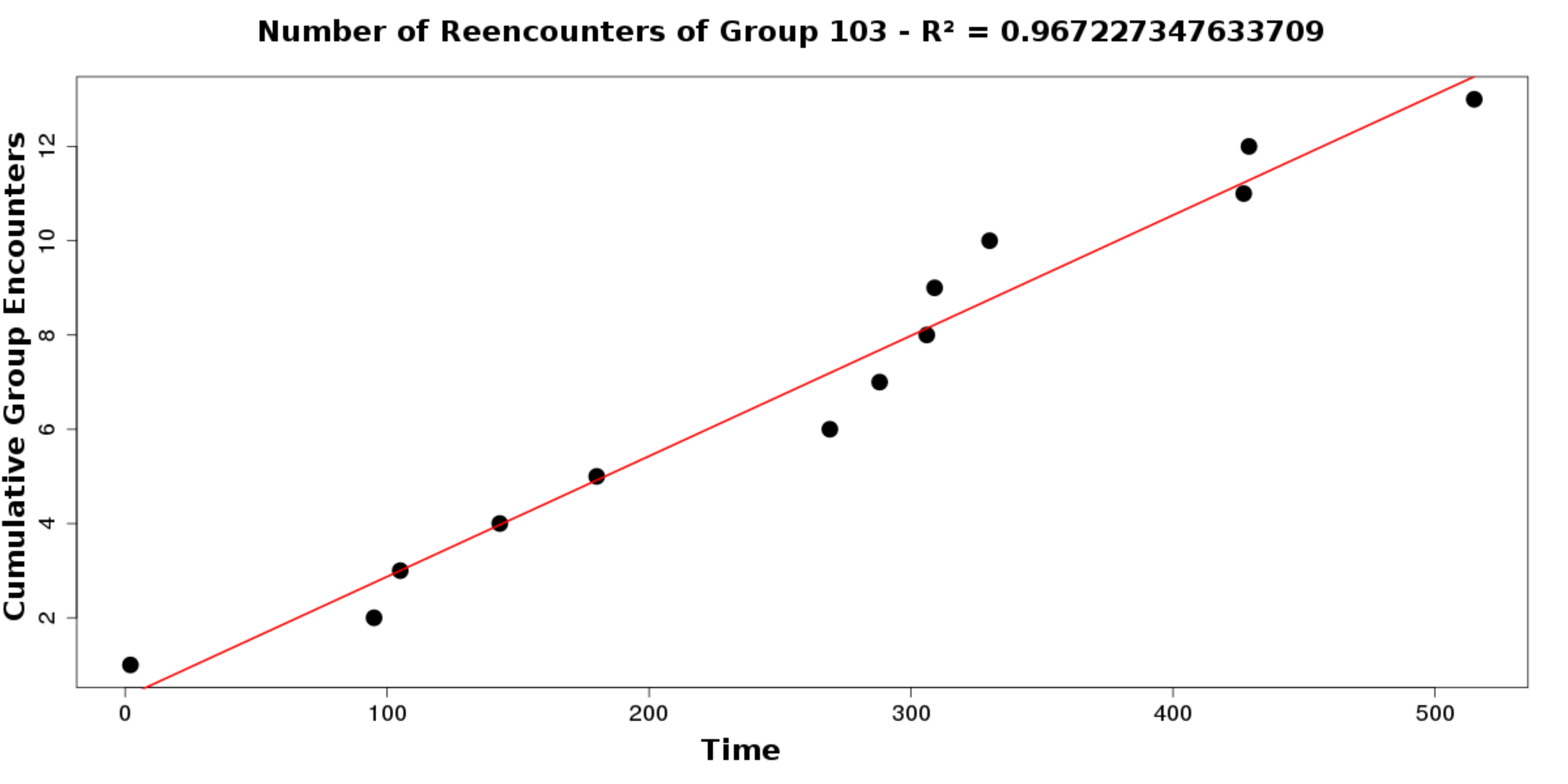}\label{p4}}
\caption{Poisson process fit for different values of $R^2$}\label{poisson}
\end{figure*}

\begin{figure}[!h]
\centering
\includegraphics[width=2.2in]{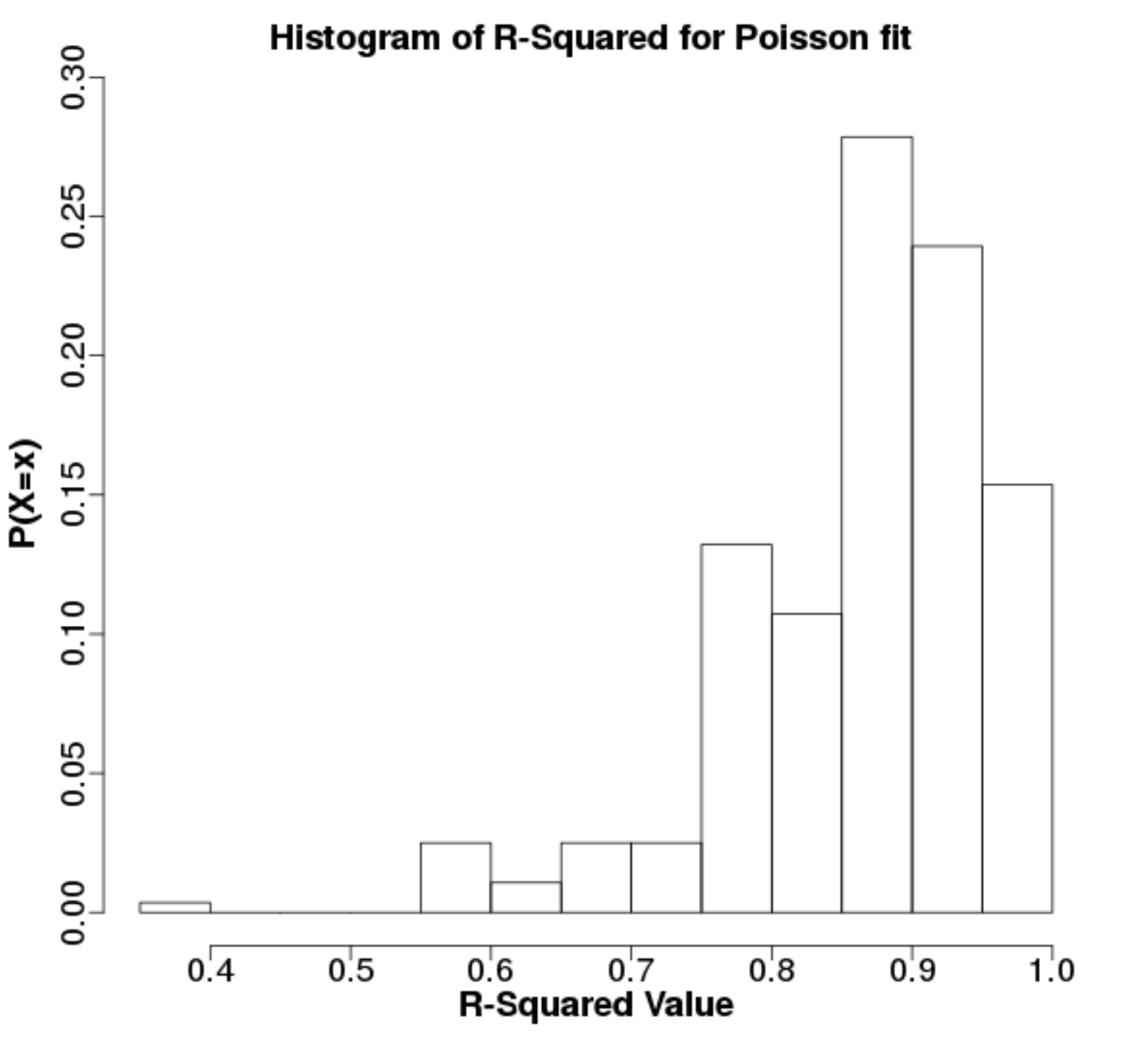}\label{b3}
\caption{R-squared distribution for Poisson distribution fits of each group of the trace}\label{R2}
\end{figure}

\change{To use group meetings in the design of a forwarding policy, there must be a representative statistical model for group meetings regularity. To model such behavior we choose a Poisson process due to its simplicity and due to the likely-hood of groups to meet again if they have met recently (result presented in Fig.~\ref{b2}). In a homogeneous Poisson process, events arrive at a constant rate $\lambda$ and the cumulative number of events over time is well approximated by a straight line with slope $\lambda$. Thus, in order to verify if group meetings can be modeled by a homogeneous Poisson process, for each group in the trace, we perform a linear regression of the number of meetings over time. Then, we compute the coefficient of determination~\cite{r2} ($R^2$) value of each group's regression, which measures how well the linear model fits to the number of meetings. Figure~\ref{poisson} exemplifies such regression for different groups, with different $R^2$ values. Figure~\ref{R2}, which presents the frequency distribution of $R^2$ for all groups in the trace, suggests that group meetings can be modeled reasonably well by a homogeneous Poisson process, since most of  $R^2$ values are 0.85 or higher. We use this Poisson process model to design our forwarding algorithm, as discussed in Section~\ref{groups}.}

\section{GROUPS-NET: Group Meetings Aware Routing}\label{groups}

Considering the group meetings properties revealed in Section~\ref{properties}, our algorithm, GROUPS-NET (Group ROUting in Pocket Switched-NETworks), works by forwarding the messages from the origin node to the destination node through the most probable group-to-group route. To model the probability of group-to-group paths, GROUPS-NET uses a probabilistic graph model in which each group detected in the recent past is represented as a node and the edges between two nodes represent the probability of a message being propagated from one group to another one. To assign a probability to an edge that links two groups, e.g., groups $A$ and $B$, GROUPS-NET considers (i) the probability of groups $A$ and $B$ meeting again in the near future, and (ii) the probability of a message being carried from group $A$ to group $B$ by a person who is a member of both groups. To compute such probability, GROUPS-NET relies on two main properties:
\begin{glist}{$\bullet$}{0}{0mm}
\topsep 0mm
\parskip 0mm
\partopsep 0mm
\parindent 0mm
\itemsep 0mm
\parsep 0mm
\item \textbf{Meetings regularity:} We assign to each group a probability of meeting again in a near future, which is based on the number of times that the group has met in the recent past. In Section~\ref{properties}, we showed that group meetings can be well approximated by a Poisson process, which can be used to estimate this probability. Moreover, recall the regularity property  depicted in Figure~\ref{b1}. The key insight is that the higher the number of meetings of a group in the recent past, the higher the probability of that group meeting again in the near future. By only considering meetings in the recent past, the meetings regularity property accounts for the social dynamism of human relationships.
\item \textbf{Shared group members:} In a group meeting, a message can be propagated for all nodes involved in the meeting. However, the message must be forwarded to the next group and so on, until it reaches a group where the destination node is member of. This group-to-group propagation is made by nodes that belong to both groups linked by an edge. If two groups have a higher number of member nodes in common, there is a greater probability for the message to be carried from group $A$ to group $B$, for instance. Thus, higher probabilities should be assigned to edges between groups that have more shared members.
\end{glist}

\vspace*{1mm}
To combine these two properties, GROUPS-NET assign edges' probabilities as the product of probabilities of groups $A$ and $B$ meeting again (meetings regularity property) weighted by the similarity in $A$ and $B$ of member compositions (shared group members property).

The $\lambda$ value of a group in the Poisson process is the inverse of the group's average inter-meeting time. Thus, given a fixed-time window size of length $L$, which is the considered time to look back in past (e.g., 3 weeks), the $\lambda$ of a group can be estimated by:
\begin{equation}
\lambda = \frac{\textit{number\_of\_meetings}}{L}.
\end{equation}

Since group meetings is well approximated by a Poisson process (as shown in Section~\ref{properties}), the probability of a given group to meet $K$ times in the $t$-time interval is given by the expression:
\begin{equation}
P[N(t) = K] = \frac{e^{-\lambda t}(\lambda t)^K}{K!}.
\end{equation}

For our opportunistic routing algorithm, we are interested in the chance of a group to meet again at least one time during the considered time interval $t$:
\begin{equation}\label{eq3}
P[N(t) \geqslant 1] = 1 - P[N(t) = 0] = 1 - e^{-\lambda t}.
\end{equation}

\change{Equation~\ref{eq3} provides a simple mechanism to compute the probability of a group to meet again at least once before a near future time $t$, given such group`s meeting frequency ($\lambda$). Conversely, GROUPS-NET sets nodes' probabilities according to Equation~\ref{eq3}. The time $t$ should be set according to the messages' TTLs, which is a parameter of the D2D network that determines the maximum tolerated delivery time for a message. In Sec.~\ref{sec:results}, we present our evaluation results as a function of TTL, because different network providers may set different TTLs for their network, depending on services, policies, and on types of applications they might want to support.}

To consider the probability of the message being propagated between two different groups by common members of both, the algorithm computes the overlap in groups' members composition as:
\begin{equation}\label{eq3.5}
P(m:G1 \rightarrow G2) = \frac{|V(G1) \cap V(G2)|}{|V(G1) \cup V(G2)|}.
\end{equation}

After setting the edges probabilities, the algorithm re-computes each edge weight as the product of each of the groups' re-meeting probabilities (computed with Equation~\ref{eq3}) multiplied by the groups' composition overlap, as in Equation~\ref{eq4}.
\begin{equation}\label{eq4}
\begin{split}
W(E_{G1,G2}) = P(G1 \rightarrow G2) \times P_{G1}[N(t) \geqslant 1] \times \\
P_{G2}[N(t) \geqslant 1]
\end{split}
\end{equation}

\change{Therefore, with edges' probabilities set, the probability of a given group-to-group route $R$ can be computed as the product of each edge in its path (Equation~\ref{eq5}). By exploiting the logarithm-likelihood property described in Equation~\ref{eq6}, the most probable group-to-group route can be computed by a shortest path algorithm, such as Dijkstra, after setting each edge weight $W(E_{i,j})$ to $-\log(W(E_{i,j}))$.}

\begin{equation}\label{eq5}
P(R) = \prod{W(E_{i,j})}, E_{i,j} \in R.
\end{equation}
\begin{equation}\label{eq6}
\begin{split}
\mbox{arg\_max}(\prod^{R}{W(E_{i,j})})&= \mbox{arg\_max}(\log(\prod^{R}{W(E_{i,j})}))\\
&= \mbox{arg\_max}(\sum^{R}{\log(W(E_{i,j})})).
\end{split}
\end{equation}

\change{Using this model, we propose GROUPS-NET to compute the most probable group-to-group path for a message to be forwarded through. During D2D network operation, each device that currently carries a copy of the message will forward such message to a newly encountered device only if the encountered device belongs to at least one of the groups in the most probable group-to-group path. Algorithm~\ref{groups_alg} presents the full picture of GROUPS-NET.}

\begin{algorithm}[]\label{groups_alg}
\SetKwInOut{Input}{inputs}
\SetKwInOut{Output}{output}
\Input{
\begin{itemize}
    \item \textbf{T}: The time window in the past to keep track of group meetings;
    \item \textbf{L}: The list of groups detected within $T$ including the number of meetings of each group $Gi$ $\in$ $L$ within period $T$;
    \item \textbf{TTL}: Messages` time to live in the network;
    \item \textbf{o}: Message origin;
    \item \textbf{d}: Message destination;
\end{itemize}
}
\Output{
\begin{itemize}
    \item \textbf{ForwardingList:} The list of devices that, if in proximity, should receive a copy of the message
\end{itemize}
}
 \ForAll{Gi $\in$ L}{
        $\lambda_i \leftarrow \frac{\mbox{Meetings}(Gi)}{T}$\\

  $P(Gi) \leftarrow 1 - e^{-\lambda_i \times TTL}$
 }
 G[V,E] $\leftarrow \emptyset$       $//$\textit{an empty graph of groups}\\
 \ForAll{\mbox{pairs} (Gi,Gj) in L}{
        $W(Gi,Gj) \leftarrow \frac{\mbox{devices}(Gi) \,\cap\, \mbox{devices}(Gj)}{\mbox{devices}(Gj) \,\cup\, \mbox{devices}(Gi)}$\\
        G[V,E].add\_edge(Gi,Gj)\\
        G.E(Gi,Gj).weight $\leftarrow$\\
        $-\log(W(Gi,Gj) \times P(Gi) \times P(Gj))$\\
 }
 R $\leftarrow$ \mbox{shortestPath}(G[V,E],o,d)\\
 ForwardingList $\leftarrow \emptyset$\\
 \ForAll{Network Devices Di}{
  \If{$Di \in R$}{
    ForwardingList.add(Di)\\
  }
 }
 \Return ForwardingList;\\
 \caption{GROUPS-NET route selection algorithm}
\end{algorithm}

The GROUPS-NET algorithm has an upper bound defined by the computation of the shortest path in a graph, which has time complexity of $O(V^2)$, where $V$ is the number of different groups in the network, i.e., vertices in the groups graph $G[V,E]$ of Algorithm~\ref{groups_alg}.

Notice that to compute the most probable group-to-group path, it is necessary to centralize the information about recent group meetings at some point. Such computation is made possible by the D2D architecture, which defines a centralized control plane and a decentralized data plane. This is the reason why GROUPS-NET properly fits applications in the D2D networks, but it is not necessarily feasible in purely distributed DTNs. To centralize the information about groups meetings, each device must periodically (e.g., once a day) update the base station with its recent group meetings.

When a given origin device wishes to send a content to a destination, it sends a request to the base station, which computes the most probable group-to-group path and sends it back to the origin device. Next, the forwarding policy proceeds as follows: starting by the origin device, each device will make the decision of forwarding or not the content to a new encountered device based on the condition that the encountered device must be a member of at least one of the groups that belong to the most probable group-to-group path.

\section{Synthetic vs Real-World Mobility}\label{sec:SRM}

\begin{figure*}[!t]
\centering
  \subfigure[MIT (Real Trace)]{\includegraphics[width=\columnwidth]{group_reencounters.pdf}\label{peri_mit}}
  \subfigure[Dartmouth (Real Trace)]{\includegraphics[width=\columnwidth]{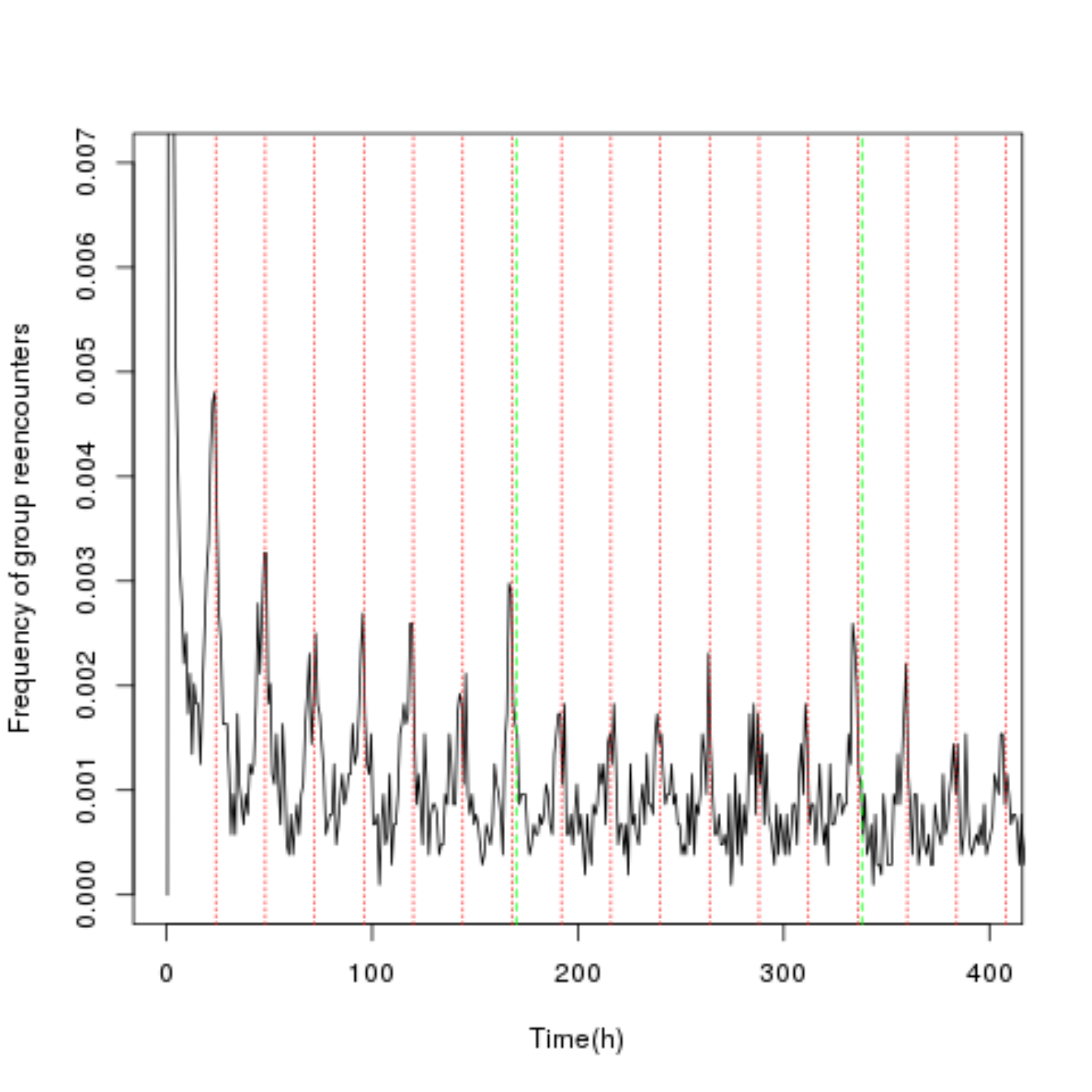}\label{peri_dartmouth}}
  \subfigure[SWIM (Synthetic Trace)]{\includegraphics[width=\columnwidth]{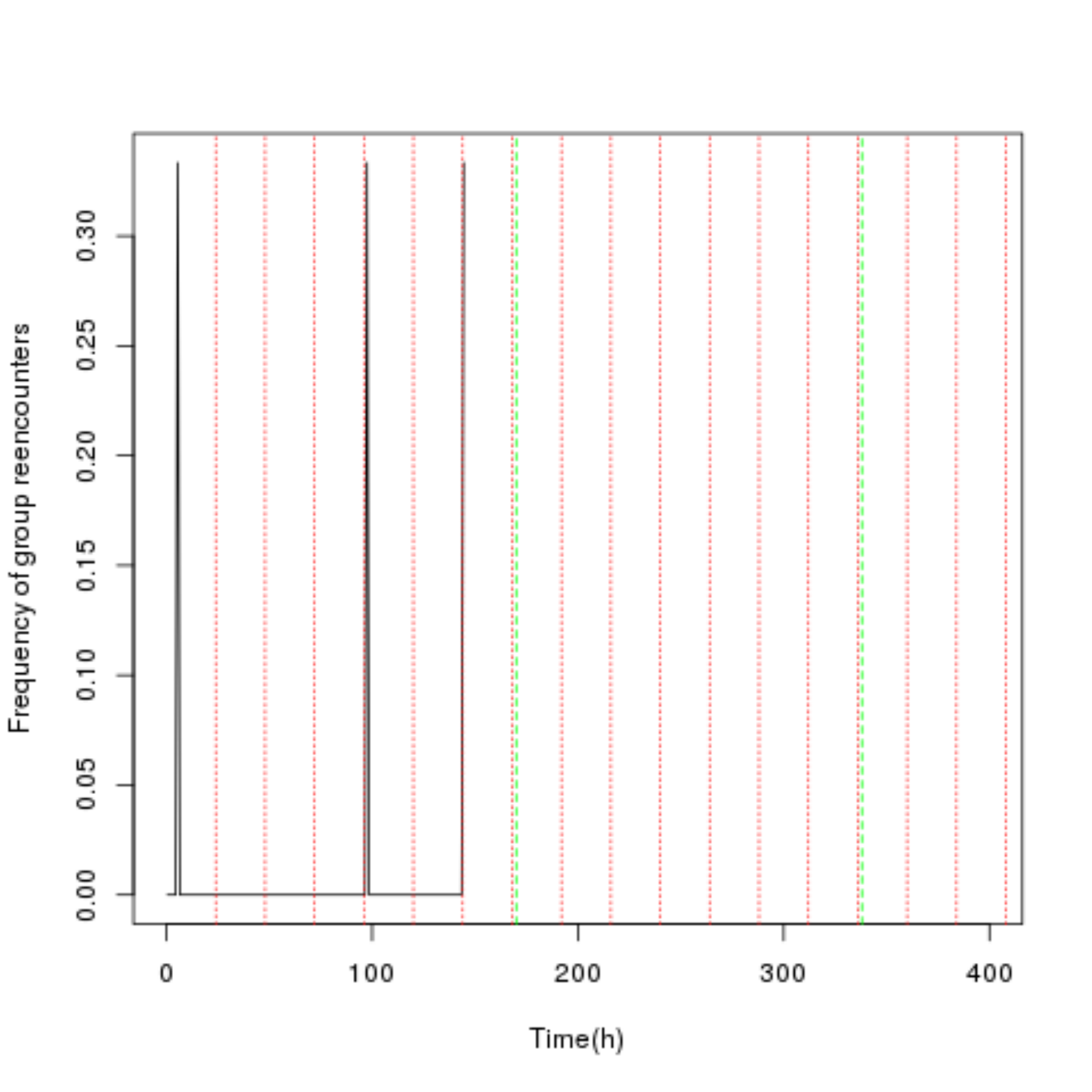}\label{peri_swim}}
  \subfigure[WDM (Synthetic Trace)]{\includegraphics[width=\columnwidth]{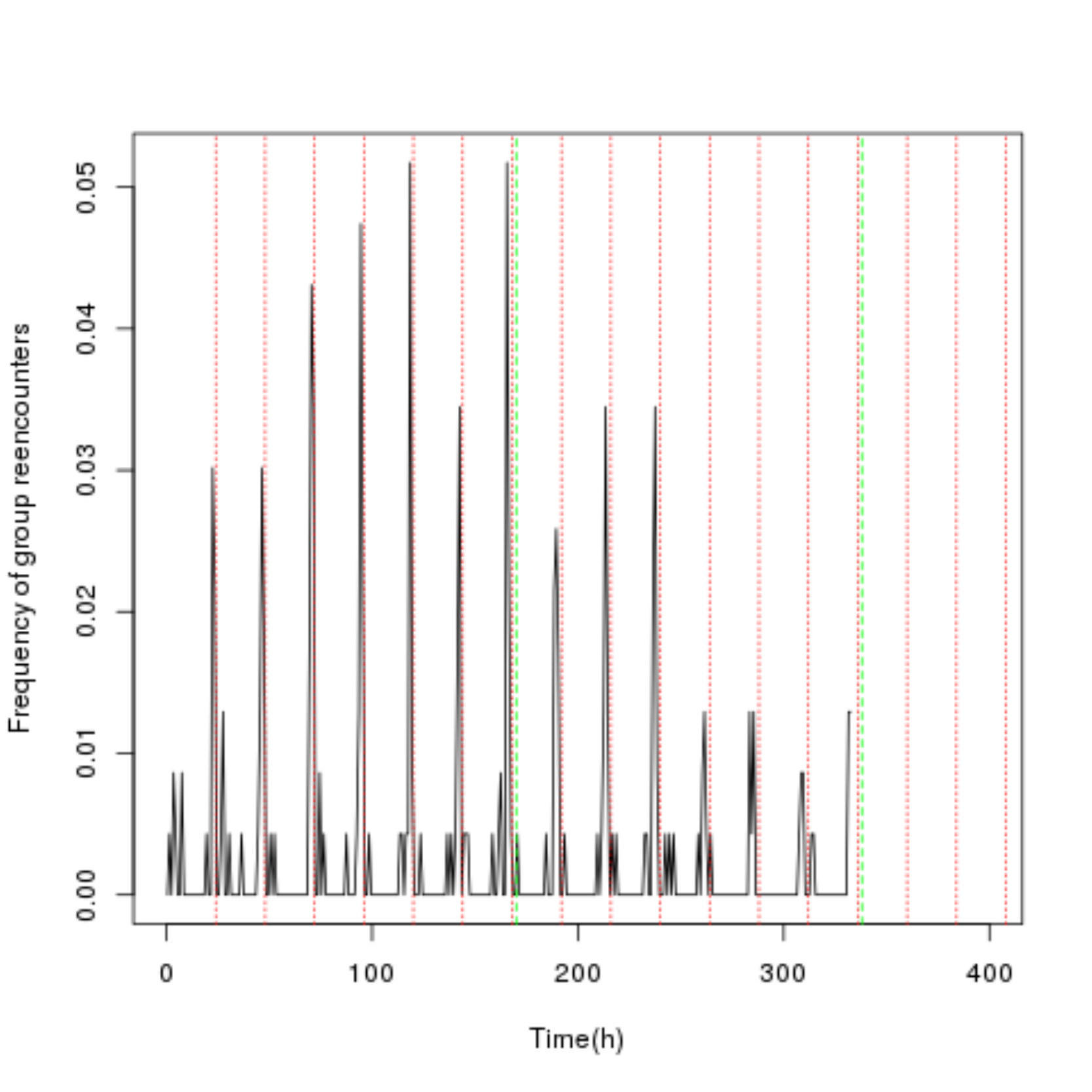}\label{peri_wdm}}
\caption{Comparison of group meetings periodicity in real and synthetic mobility traces}\label{periodicity}
\end{figure*}

Before evaluating GROUPS-NET, we verify if we could use synthetic traces to validate it. To do that, we compare state-of-the-art synthetic mobility models with real mobility traces, with the goal of verifying if group meetings' regularity properties are captured by such synthetic models. Specifically, we study if such models capture group re-encounters and their evolution over time to decide if they are representative, considering the group mobility feature and if they should or not be used in the validation of opportunistic networking protocols based on group meetings and social context.

\subsection{Real-World Mobility Models}

We apply the methodology for detecting and tracking groups, defined in Section~\ref{sec:GMD}, to two real mobility traces: MIT and Dartmouth. Figures~\ref{peri_mit} and~\ref{peri_dartmouth} show the P.D.F.\ (Probability Density Function) of group re-meetings along the time for these traces. In both of them, we can verify the presence of periodicity in groups' re-encounters. Moreover, we see that the mass of probability is concentrated in peaks around the red dotted lines, which represent periods of 24 hours. We also observe in Figures~\ref{peri_mit} and~\ref{peri_dartmouth} that higher peaks are present around the green dashed lines, which represent periods of seven days. As discussed in Section~\ref{properties}, this pattern in the group re-meetings' P.D.F.\ shows that group meetings present daily and weekly periodicity. It is noteworthy that such pattern is observed in both real traces, even though they are from different places, have different number of nodes, and used different data collection methods. Next, we analyze three widely used state-of-the-art synthetic mobility models to verify if they properly represent social groups' regularity properties.

\subsection{Synthetic Mobility Models}

The SWIM mobility model~\cite{swim} generates synthetic small worlds, preserving the pairwise contact duration and inter-contact times statistical distributions as observed in real mobility traces. The SLAW mobility model~\cite{slaw} captures several significant statistical patterns of human mobility, including truncated power-law distributions of human displacements, pause-times and pairwise inter-contact times, fractal way-points, and heterogeneous areas of individual mobility. The Working Day Movement (WDM) synthetic model~\cite{wdm} also captures the same statistical properties of contact durations and inter-contact times as found in SWIM and SLAW. However, in addition to these properties, WDM captures the daily regularity of human movements.

As we did for the real traces, MIT and Dartmouth, we have applied our group detection and tracking methodology to the contact traces generated by these three synthetic models. Figures~\ref{peri_swim} and~\ref{peri_wdm} present the results for the SWIM and WDM models, respectively.

The contact trace generated by the SWIM model (Figure~\ref{peri_swim}) do not present any regularity in group meetings. Out of the detected groups only three group re-meetings were registered in a period of 15 days. The result for the contact trace generated by the SLAW model presented an analogous behavior, i.e., no regularity in group meetings. This behavior is explained by the fact that such models were designed to be representative of the statistical properties of pairwise contacts only, without considering that human contacts often involve more than two peers. These models look only at pairwise contacts, disregarding group meetings.

In the WDM trace (Figure~\ref{peri_wdm}) we can observe that group re-meetings happen precisely in periods of 24 hours and with much higher frequencies than in real mobility traces. This behavior is observed because WDM initially defines a set of places, called offices, and then distributes nodes to transition between pre-defined subsets of offices with daily periodicity. Therefore, nodes with intersections in their lists of offices will always form groups with an exaggerated meeting regularity.

By analyzing the group meetings regularity of the synthetic models, we conclude that none of them can accurately represent group mobility patterns. For this reason, none of the synthetic models is suitable for evaluating GROUPS-NET, which is based on the group meetings' regularity. Therefore, in Section~\ref{sec:results}, we evaluate GROUPS-NET using only real mobility traces, which do not suffer from such biases. We also highlight the need for designing mobility models that better represent the role of social groups and their regularity in human mobility.

\section{Comparative Analysis}\label{sec:results}

To validate the performance of GROUPS-NET, we compare it to the forwarding algorithm that achieved the most cost-effective performance in D2D networks: Bubble Rap~\cite{bubble}.

\subsection{Bubble Rap Algorithm}

The Bubble Rap algorithm identifies static social communities by looking at densely interconnected nodes in the aggregated contact graph in the whole trace using the Clique Percolation Method~\cite{palla2005}. Therefore, each node in the network must belong to at least one community. Nodes that do not belong to any community are assigned to a pseudo-community of one node. This is necessary in the forwarding algorithm operation. Moreover, each node gets a measure of its global popularity in the network (\textit{GlobalRank}) and a local measurement of popularity, which is valid within that node's community (\textit{LocalRank}). Using these parameters, the forwarding strategy works as follows:
\begin{glist}{$\bullet$}{0}{0mm}
\topsep 0mm
\parskip 0mm
\partopsep 0mm
\parindent 0mm
\itemsep 0mm
\parsep 0mm
\item At each encounter, a given node transmits its content if the encountered node has a higher \textit{GlobalRank}, or if the encountered node belongs to a community of which the final destination is a member.

\item Once the message is inside the final destination's community, the forwarding process occurs if the \textit{LocalRank\/} of the encountered node is higher than the \textit{LocalRank} of the node that has the message. This procedure goes on until the message reaches the final destination.
\end{glist}

\vspace*{1mm}
With the purpose of having a fair comparison, in this work we implemented Bubble Rap using the community detection pre-calibrated parameters reported in~\cite{bubble}, which provided the best results in terms of cost-effective content delivery. Also, the \textit{GlobalRank\/} and the \textit{LocalRank\/} were calculated using the C-Window technique that better approximated the node centrality metric in their experiments~\cite{bubble}.

\subsection{Performance Evaluation}

\begin{figure*}[!t]
\centering
  \subfigure[Delivery (MIT)]{\includegraphics[width=2.8in]{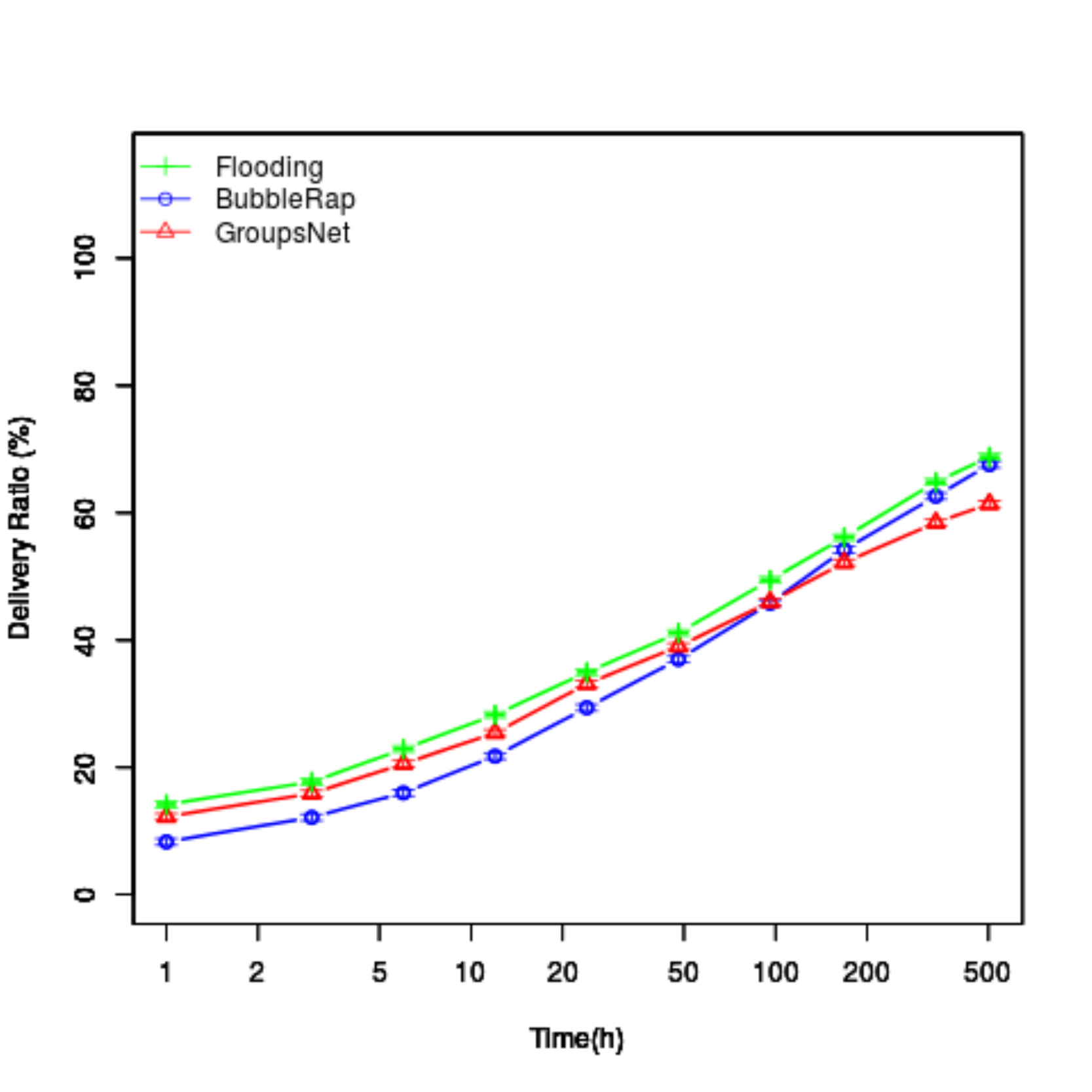}\label{g1}}
  \subfigure[Transmissions (MIT)]{\includegraphics[width=2.8in]{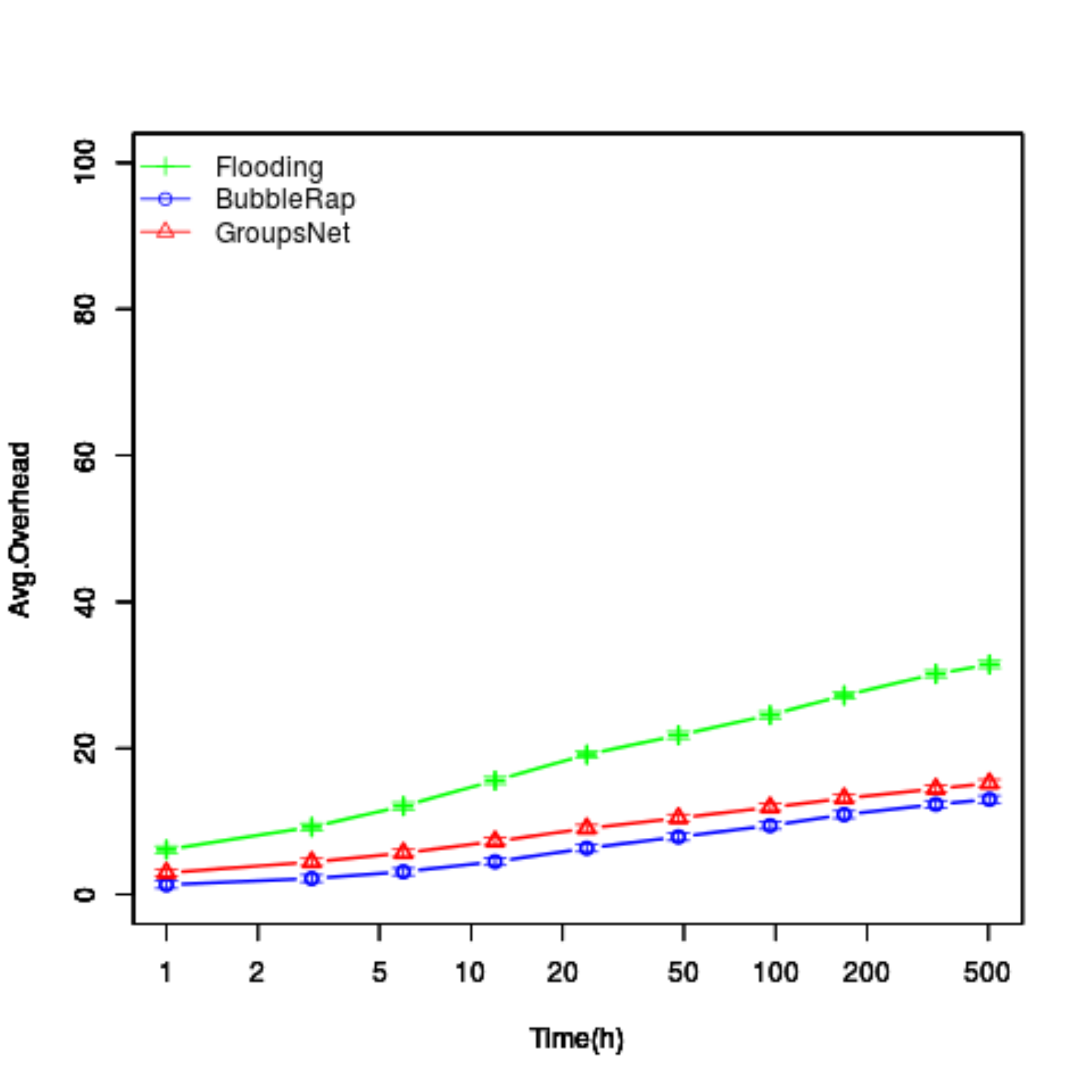}\label{g2}}
  \subfigure[Delivery (Dartmouth)]{\includegraphics[width=2.8in]{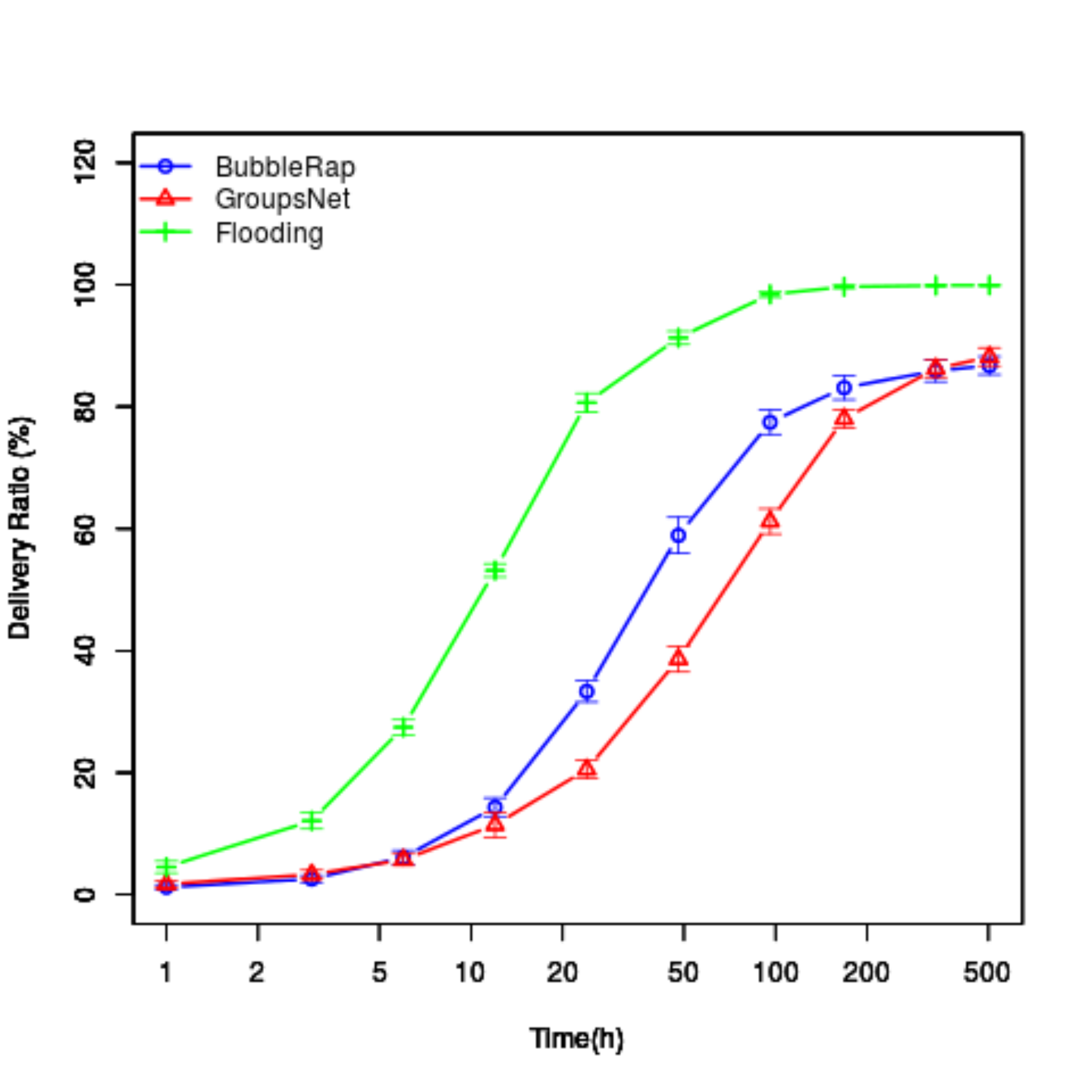}\label{g3}}
  \subfigure[Transmissions (Dartmouth)]{\includegraphics[width=2.8in]{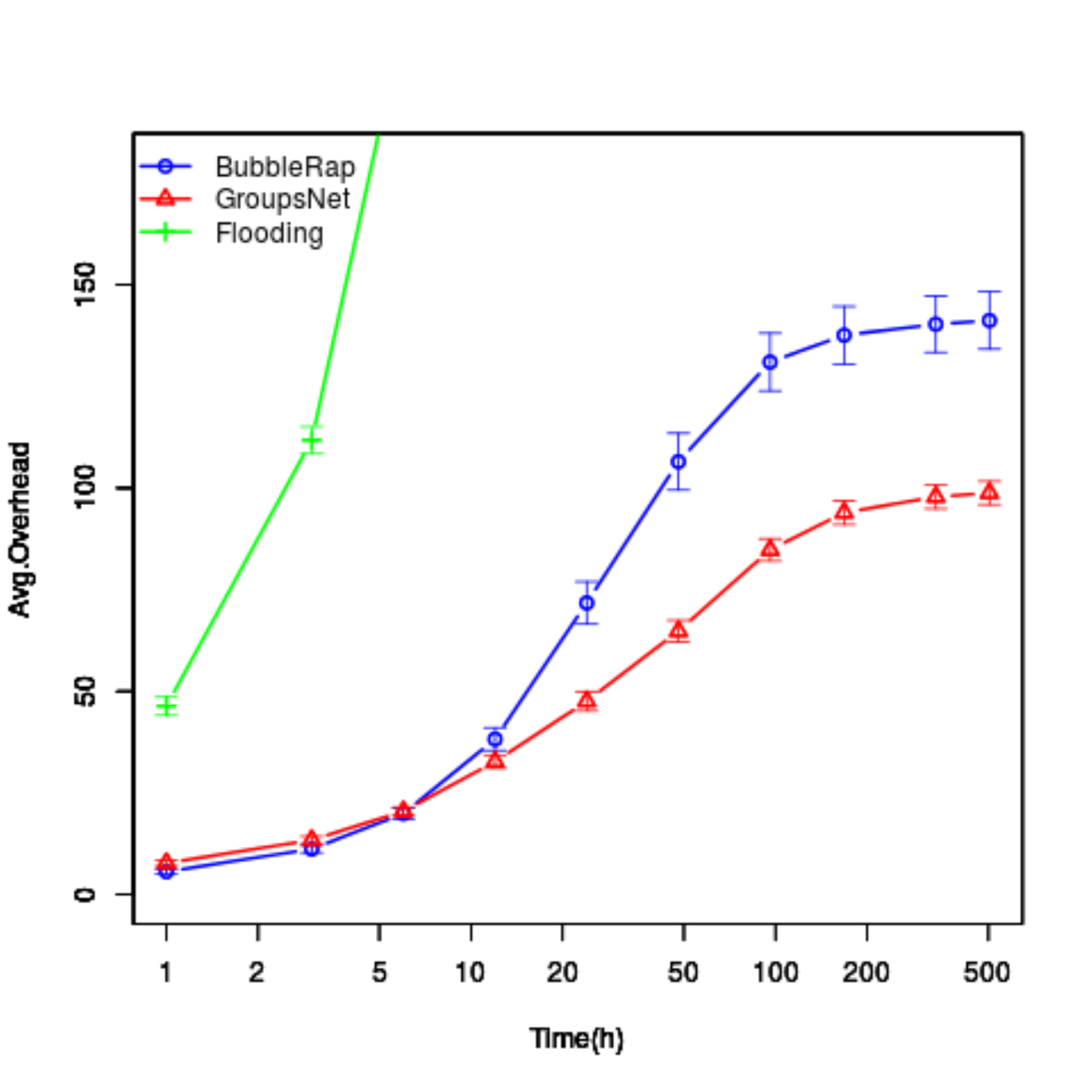}\label{g4}}
\caption{Delivery ratio and network overhead of Bubble Rap and GROUPS-NET}\label{result}
\end{figure*}

\begin{figure*}[!t]
\centering
  \subfigure[Benefit-cost in log time scale (Dartmouth)]{\includegraphics[width=2.8in]{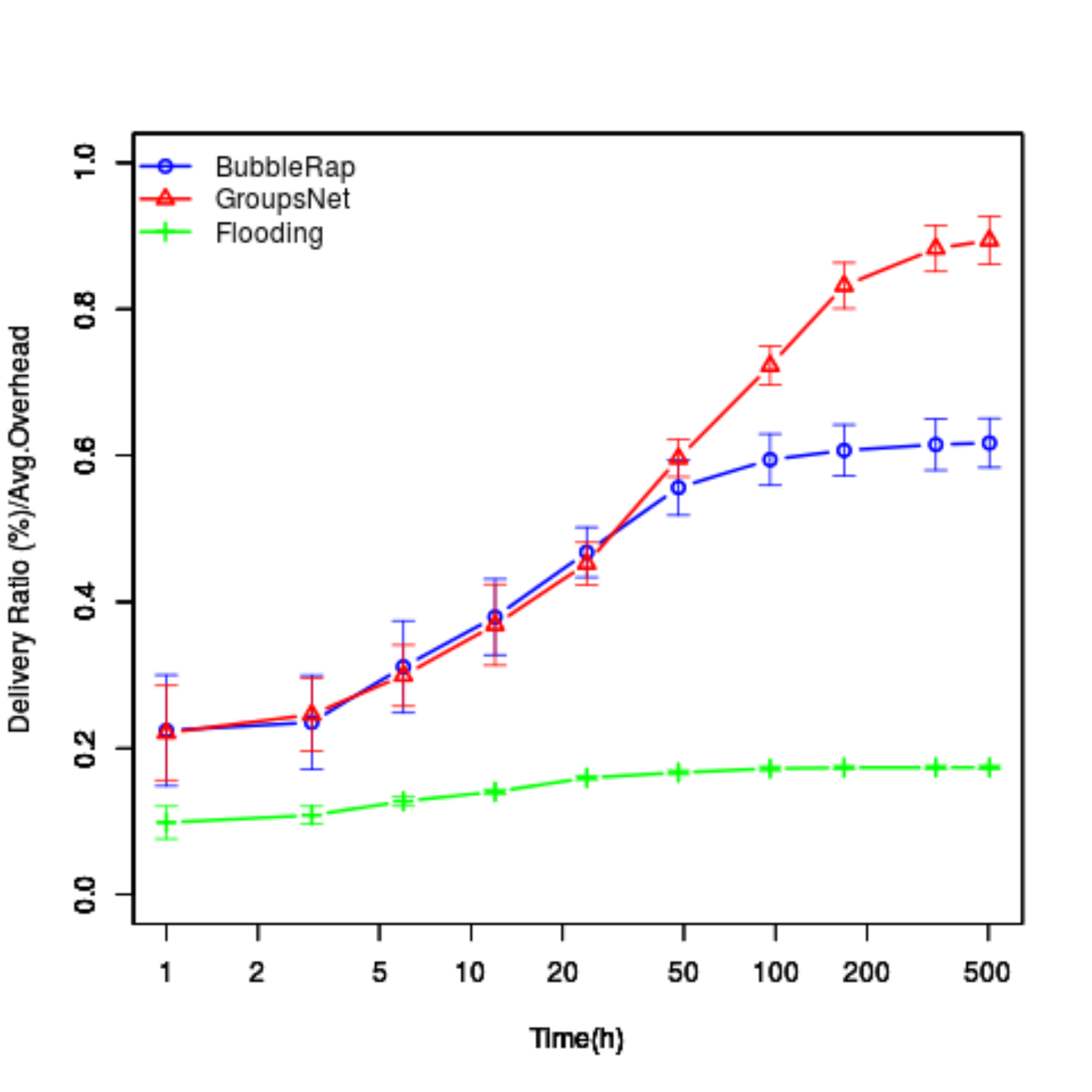}\label{g5}}
  \subfigure[Benefit-cost in regular time scale (Dartmouth)]{\includegraphics[width=2.8in]{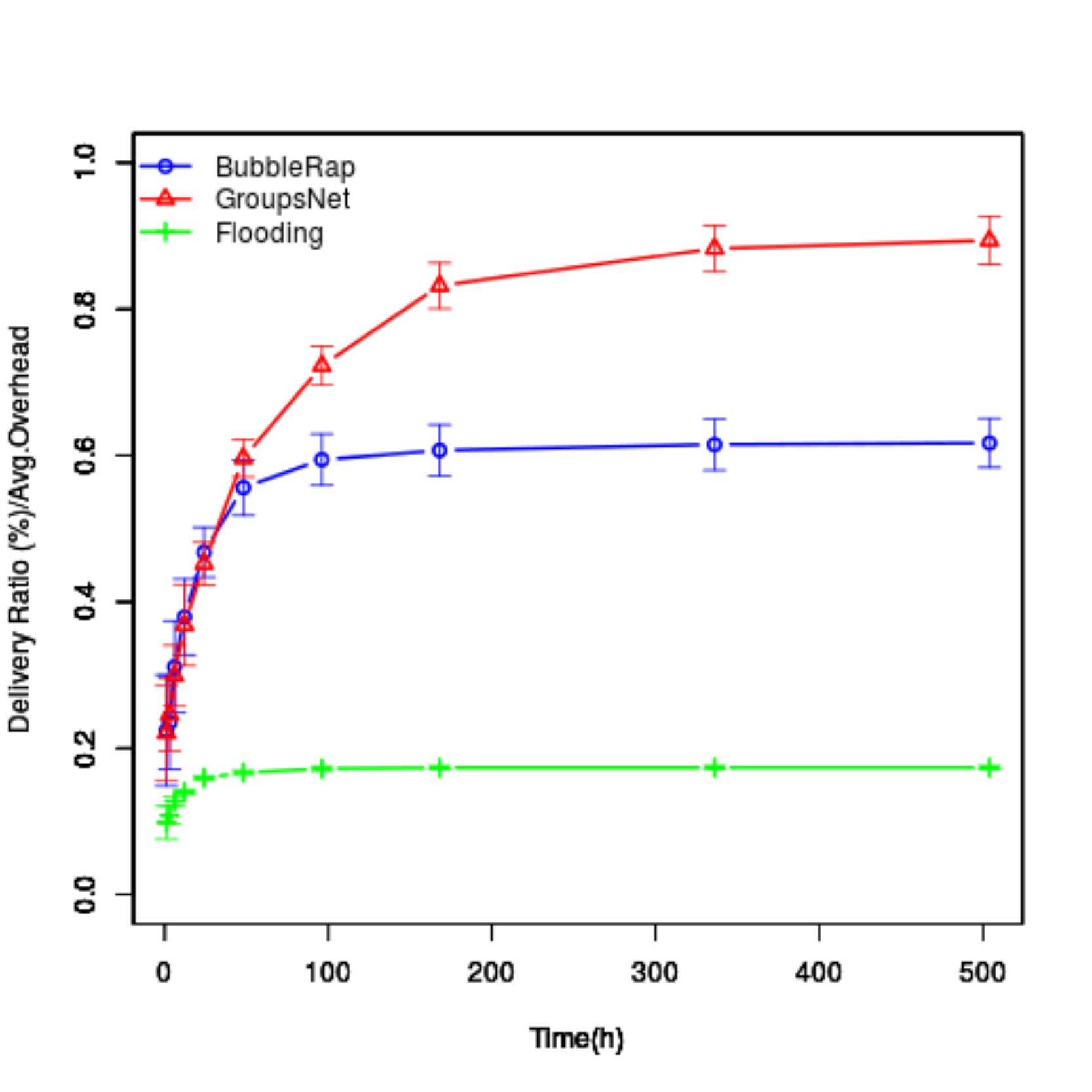}\label{g6}}
\caption{Cost-effectiveness comparison of Bubble Rap and GROUPS-NET}\label{result2}
\end{figure*}

We used the following metrics in the evaluation of GROUPS-NET and Bubble Rap:
\begin{glist}{$\bullet$}{0}{0mm}
\topsep 0mm
\parskip 0mm
\partopsep 0mm
\parindent 0mm
\itemsep 0mm
\parsep 0mm
\item \textbf{Delivery ratio:} evaluates the percentage of successfully delivered messages along the time;
\item \textbf{Number of transmissions:} measures the network overhead, i.e., the number of D2D transmissions that each algorithm performs along the time.
\end{glist}

For each evaluated trace (MIT and Dartmouth), an (\textit{origin, destination\/}) pair is randomly selected among the users of the trace with uniform probability. Moreover, the time $t$ in which the message transmission starts at the origin node is also randomly selected within the trace duration. Both protocols were executed with 500 randomly generated pairs (\textit{origin, destination, time\/}). This process was repeated eight times with different seeds for random number generation, to obtain 95\% confidence intervals.

This way, we aim to capture varied behavior patterns throughout the trace, conferring generality to the tests. Together with GROUPS-NET and Bubble Rap results, we also plot the results for a flooding transmission. In the flooding scheme, the message is always propagated whenever a node that has a message encounters a node that does not have it yet. The flooding establishes the upper bound for the delivery ratio and for the network overhead. In our tests, the recent past period used by GROUPS-NET to predict future group meetings was of three weeks, since it is enough to capture both daily and weekly periodicities.

Figure~\ref{result} presents the comparative results in terms of delivery ratio and network transmissions overhead along the time. The results for the MIT Reality Mining trace, presented in Figures~\ref{g1} and~\ref{g2}, show that in the first hours, after the beginning of a transmission, GROUPS-NET has a slightly higher delivery ratio and, in the final hours, Bubble Rap overcomes it, successfully delivering a small higher percentage. Throughout the whole message propagation time, GROUPS-NET presented a slightly higher network overhead. This happens because the MIT Reality Mining trace is very particular, since all monitored users reside in the same university buildings. For this reason, they are expected to have stronger social bonds than regular nodes in D2D scenarios. This characteristic benefits an algorithm such as Bubble Rap, which uses the static social structure in its forwarding policy. This fact motivated the study of a large scale and more general trace, such as Dartmouth, to evaluate the forwarding tests. However, notice that even in this specific scenario, GROUPS-NET presented a competitive result when compared to Bubble Rap, with the advantage of not requiring community detection and parameters' calibration, which are hard or unfeasible in a practical real-time scenario.

Figures~\ref{g3} and~\ref{g4} present the results for the same experiment performed in the Dartmouth trace, which is more general and has a larger scale. In this scenario, GROUPS-NET achieved a considerable better performance than Bubble Rap. In the period from 24 to 96 hours after the start of the message propagation, Bubble Rap obtains a higher delivery ratio but, after that, it is outperformed by GROUPS-NET until the end of the three weeks' transmission period. With respect to the network overhead, after the sixth hour, Bubble Rap starts to transmit much more messages than GROUPS-NET, presenting an average overhead 50\% higher in the following hours. For 1000 different (\textit{origin, destination\/}) messages, this represents an economy of 60000 D2D-transmissions in the network.

Here we define a benefit-cost metric as the ratio between successfully delivery and network overhead. Throughout the duration of the transmissions, GROUPS-NET presents a better benefit-cost than Bubble Rap. As depicted in Figures~\ref{g5} (in log-scale) and~\ref{g6} (in regular-scale), after the first hours of transmission, GROUPS-NET reaches two times Bubble Rap's benefit-cost. As mentioned before, GROUPS-NET also has the advantage of not depending on community detection schemes nor needing parameter calibration, being for these reasons a viable practical solution.

\subsection{Discussion}

\begin{figure*}[!t]
\centering
  \subfigure[]{\includegraphics[width=2.2in]{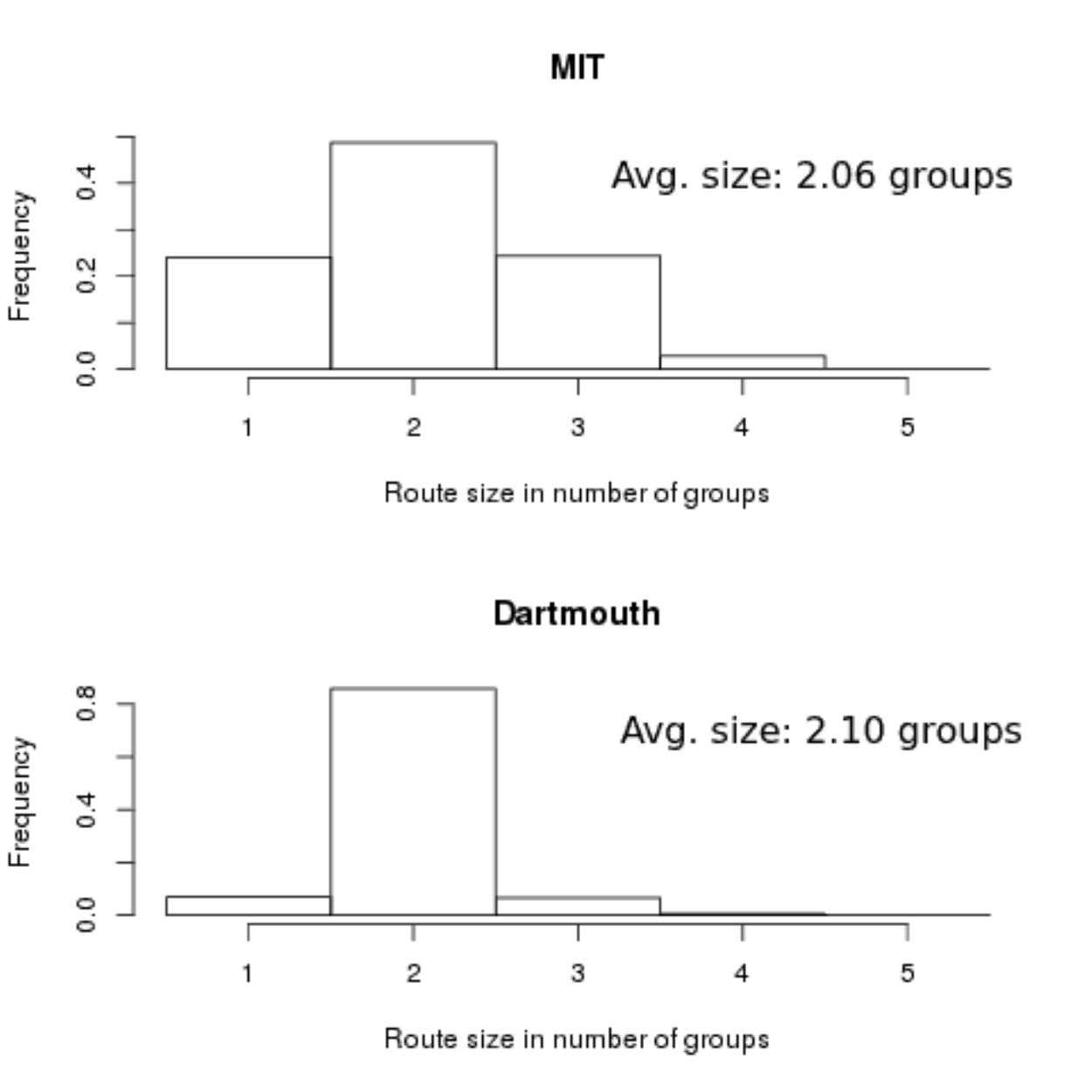}\label{tam_rotas}}
  \subfigure[]{\includegraphics[width=2.2in]{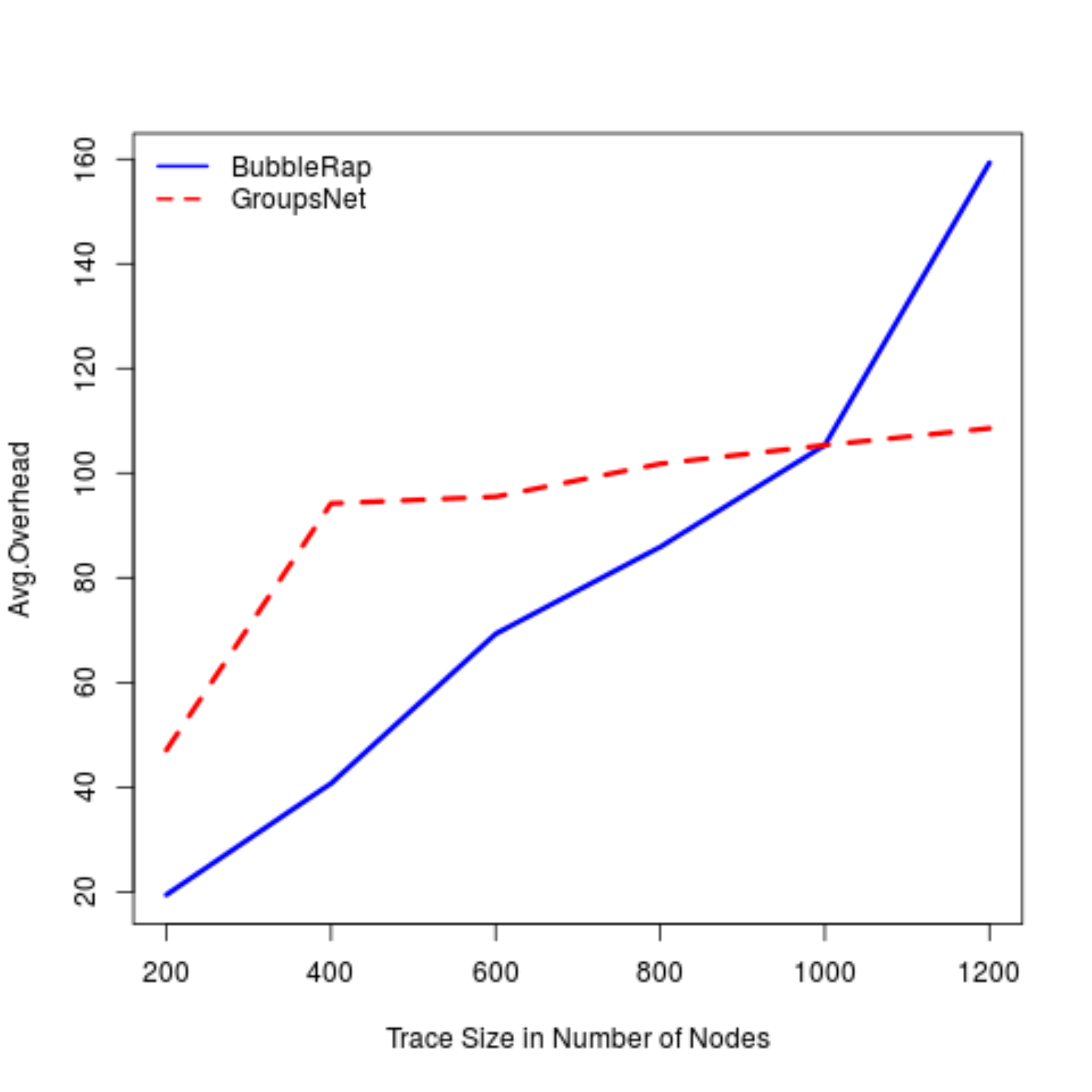}\label{medias_overhead}}
  \subfigure[]{\includegraphics[width=2.2in]{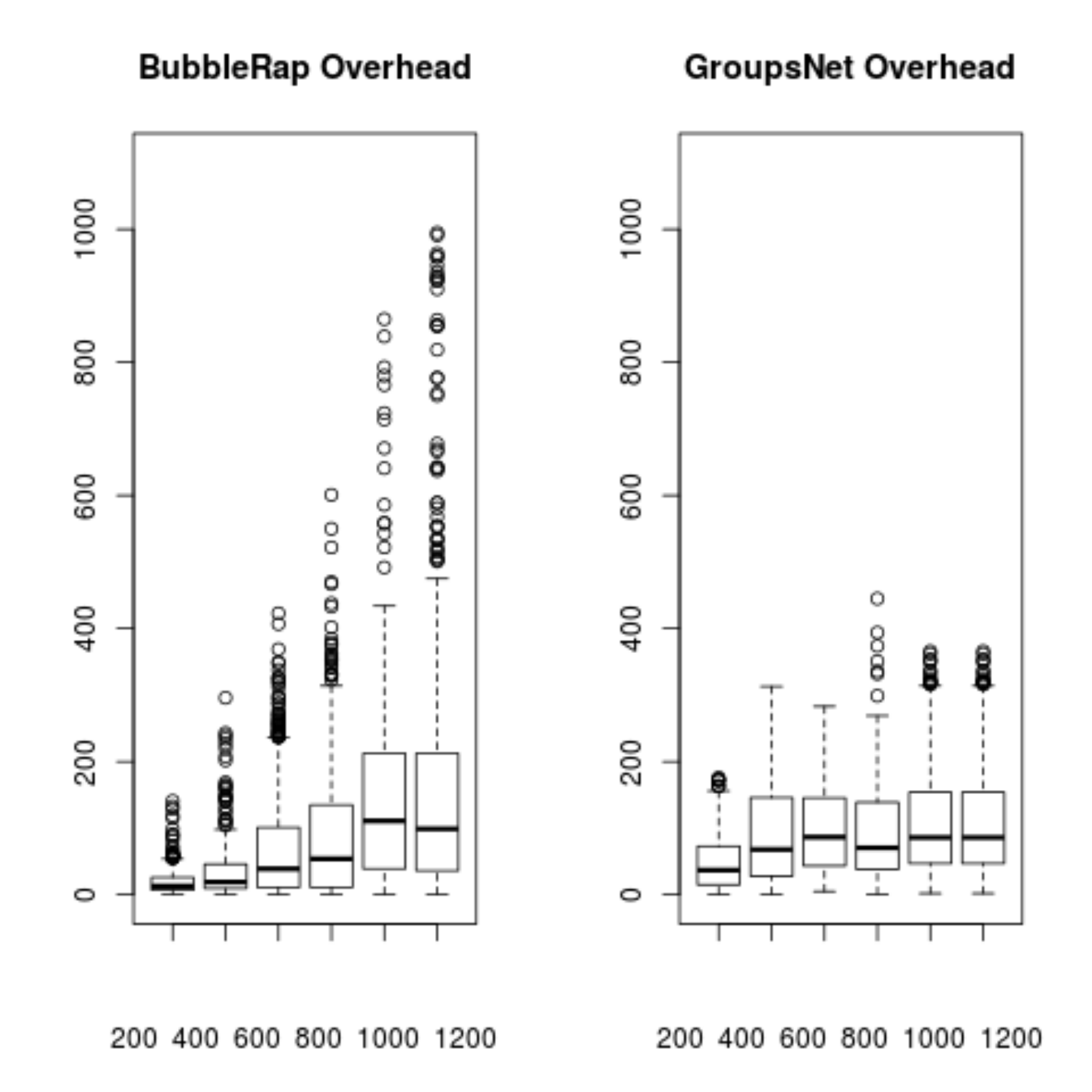}\label{overhead_evos}}
\caption{(a) Distribution of sizes of the most probable group-to-group paths in the MIT Reality Mining and Dartmouth traces; (b) Average network overhead for different scale scenarios; (c) Boxplots of the network overhead for 1000 transmissions in different trace scales}\label{overhead}
\end{figure*}

By forwarding messages through the most probable group-to-group path, GROUPS-NET achieves a high delivery ratio, which is comparable to the upper bound of flooding and to Bubble Rap in both small-scale and large-scale scenarios. This conforms the result presented in Figure \ref{b2}. However, when looking at Figure~\ref{result}, one question that arises is: why does GROUPS-NET present a much lower overhead in large-scale scenarios, but not in small-scale ones? The answer to this question rests in the nature of each algorithm.

Bubble Rap works by forwarding messages to nodes that have a higher popularity. Therefore, the maximum number of transmissions is limited by the nodes in the network that have higher popularity than the origin. When the origin is randomly selected, the expected number of nodes that are more popular than the origin is directly proportional to the total number of nodes in the trace. For this reason, when the number of nodes in the network is increased 15 times, from 80 (in MIT) to 1200 (in Dartmouth) Bubble Rap's overhead also increases by approximately 15 times, as presented in Figures~\ref{g2} and~\ref{g4}.

GROUPS-NET maximum overhead, on the other hand, is limited by the size of the most probable group-to-group path (in number of groups) multiplied by the average number of members in the groups of such path. Figure~\ref{tam_rotas} shows the distribution of (for most probable group-to-group) path sizes in terms of the number of groups involved in the path. In both traces, for randomly selected (\textit{origin, destination\/}) pairs, the distribution of the paths' sizes is not proportional to the number of nodes involved in the network. In fact, the average size does not change for different scales. Since the addition of each new group to the route involves a multiplication by the probability of a new edge (which can significantly reduce a path probability), the most probable paths tend to have few hops. For this reason, GROUPS-NET presents significantly lower overhead in large scales.

To investigate how the network overhead evolves with the increase of the number of nodes in the network, we generate subsets of 200, 400, 600, 800 and 1000 nodes from the original Dartmouth dataset. These subsets are generated using a Snowball algorithm, which firstly assembles a social-contact graph and selects the node with the highest centrality. Next, it adds to the subset of nodes the neighbors of the central node. Then, it adds the neighbors of such neighbors and so on, until the desired number of nodes for the particular subset is reached. This way, the social structure of the network is preserved even for small subsets of 200 and 400 nodes, considering a total of 1200 nodes.

Using these subsets, we evaluated the overhead of both algorithms with the goal of analyzing their evolution with the increase of the number of nodes in the network. Figure~\ref{medias_overhead} shows the average overhead per message with different network sizes for both algorithms. Figure~\ref{overhead_evos} presents the statistical distribution of such messages' overheads for 1000 transmissions in each scenario. These experiments confirm the expected behavior for the overhead. Bubble Rap's overhead presents a linear increase with the size of the network. GROUPS-NET's overhead, on the other hand, remains stable for networks with 400 nodes or more. Since real cellular networks often have thousands (or even millions) of nodes, we claim that GROUPS-NET is better suited for such applications.

\subsection{\change{Experimental Limitations}}

\change{A recurrent problem in the validation of mobile networking protocols is the lack of real world data-sources. This is because real world data-sources are hard and expensive to collect. To mitigate this problem, synthetic mobility models are constantly proposed and improved. However, as we evaluate in section 6, state-of-art synthetic models are not representative of group meetings regularity and, therefore, our validation cannot use them. We here emphasize that it would be interesting and useful to incorporate group meeting regularity in synthetic models, since it is present in real-world traces.}

\change{Knowing that synthetic models are not suitable for our evaluation, we validated GROUPS-NET using the two largest scale publicly available data-sources. MIT Reality has the longest time duration, while Dartmouth has the highest number of monitored users. Both of them were collected inside university campuses. It is important to discuss, however, that it would be interesting to extend this evaluation to larger scales and also to datasets from other types of environments, such as a whole city. This would enable a more reliable evaluation of mobile and D2D communication protocols scalability. Such experiments will be possible when i) more mobility and contact real-world datasets with larger scale become publicly available, and ii) synthetic mobility models evolve towards generating even more realistic synthetic traces.}

\change{As mobility models evolve and more realistic datasets become available, we expect other issues that were not addressed in this work to be investigated. Examples are the evaluation of physical communication issues, such as interference for D2D link establishment and management, and device hand-off between different base stations. From another perspective, it is also important to look at the D2D communication problem from the user perspective. In this regard, device memory/energy management and user incentive mechanisms must be proposed to convince users to participate in such a collaborative network. We consider all this issues to be interesting and promising future work.}

\section{Conclusion and Future Work}\label{sec:conslusion}

In this work, we introduce the use of social group meetings awareness to leverage cost-effective message transmissions in multi-hop D2D Networks. First, we propose a methodology for detecting group meetings from contact traces. Using the detected groups, we build a probabilistic graph, which is used to compute the most probable group-to-group path for a message to be forwarded. Our approach has the advantage of not requiring community detection and of being parameter-calibration free. Our experiments show that, in large-scale scenarios, this strategy is more cost-effective than previous state-of-the-art strategy (which is based on static social communities) with respect to delivery ratio and network overhead.

The results show that the group meetings approach is a promising strategy. Based on this idea, one can propose forwarding strategies for several different applications. For example, GROUPS-NET can be modified to consider different types of forwarding such as \textit{Single-Source-Multiple-Destinations\/} or \textit{Multiple-Source-Multiple-Destinations}. This could be achieved by computing the union of all pairwise most probable group-to-group paths in the groups graph. Examples of applications that could benefit from \textit{Single-Source-Multiple-Destinations\/} and \textit{Multiple-Source-Multiple-Destinations} are those with high download demand and in which timely delivery is not essential, such as smartphones' system updates or video advertisement.

In the GROUPS-NET strategy, described in Section~\ref{groups}, only the most probable path is considered for forwarding the message. However, more than one path can be considered for forwarding a message. An interesting future work would be to propose an overhead constrained version of GROUPS-NET, which would add a \textit{maximum-overhead\/} parameter to the algorithm. This way, GROUPS-NET would forward a message through the $N$ most probable redundant group-to-group paths that involve at most the maximum tolerated number of nodes. The higher the \textit{maximum-overhead\/} parameter is, the more similar to a flooding the forwarding will behave, since the flooding strategy forwards messages through all possible paths. In the case that the single most probable path involves more nodes than the maximum tolerated overhead, then the most probable path with less nodes than the maximum overhead would be chosen instead. This strategy would allow to decrease the base station bandwidth demand and, at the same time, control the D2D network overhead, a feature that is not possible in previous forwarding strategies.

\change{In addition to the aforementioned results, we show in Section~\ref{sec:SRM} that an interesting open issue is how to correctly model the role of social group meetings in the human mobility in order to synthesize better artificial mobility traces. Finally, this work reveals the need for evaluating D2D algorithms' scalability, as previous state-of-the-art solutions seem to not fit well to large-scale networks.}

\newpage

\begin{IEEEbiography}[{\includegraphics[width=1in,height=1.25in,clip,keepaspectratio]{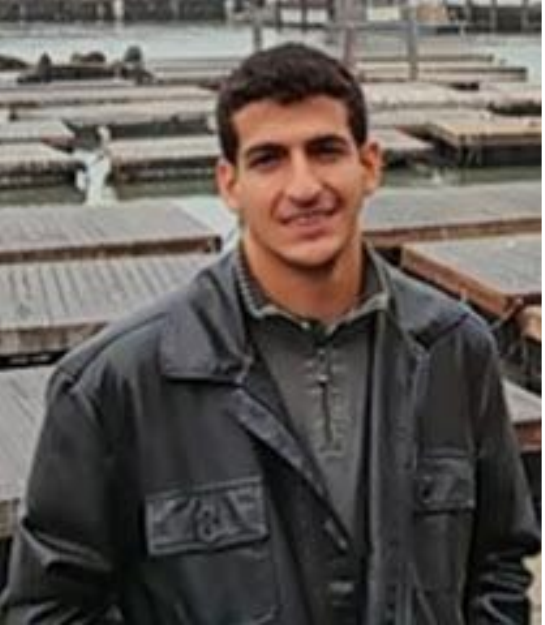}}]{Ivan Oliveira Nunes}[ivanolive@dcc.ufmg.br] is currently a networked systems Ph.D. student at University of California Irvine (UCI). He received his computer science M.Sc. from Federal University of Minas Gerais (UFMG), Brazil, in 2016, and his bachelor degree in Computer Engineering from the Federal University of Espirito Santo (UFES), Brazil, in 2014. His current research interests include networking, mobile and ubiquitous computing, security, and embedded systems.
\end{IEEEbiography}
\vspace{-1cm}

\begin{IEEEbiography}[{\includegraphics[width=1in,height=1.25in,clip,keepaspectratio]{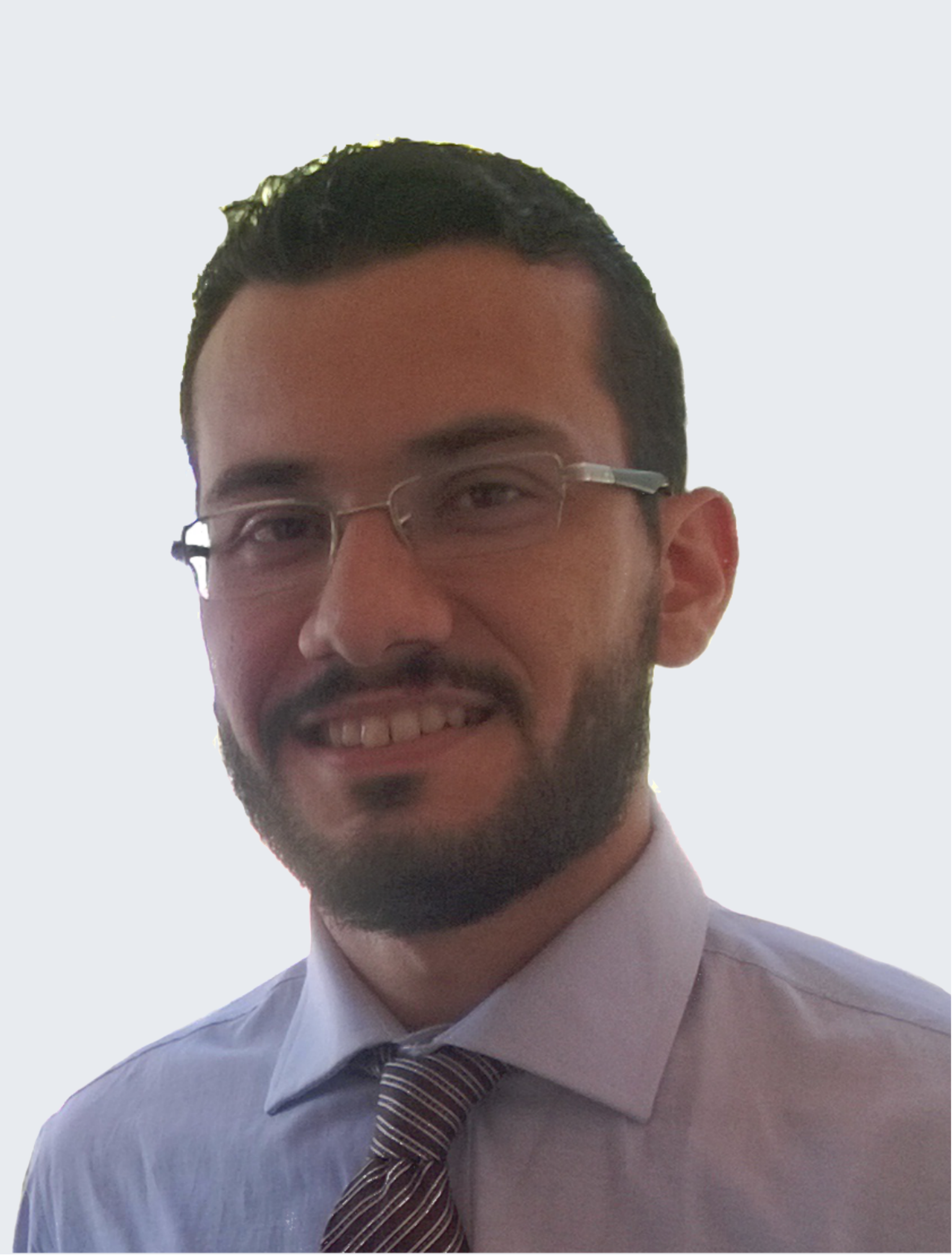}}]{Clayson Celes} is currently a computer science Ph.D. candidate at Federal University of Minas Gerais (UFMG), Brazil. He received his M.Sc. degree in Computer Science from UFMG in 2013 and his bachelor degree in Computer Science from the State University of Ceara (UECE), Brazil, in 2010. His research areas are mobile computing, vehicular networks, and ubiquitous computing.
\end{IEEEbiography}
\vspace{-1cm}

\begin{IEEEbiography}[{\includegraphics[width=1in,height=1.25in,clip,keepaspectratio]{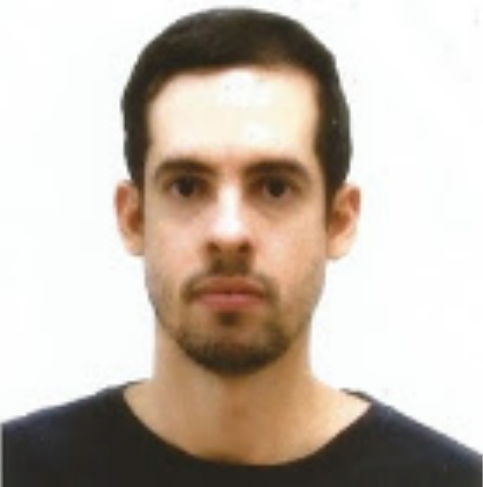}}]{Pedro O.S. Vaz de Melo} is an assistant professor in the Computer Science Department (DCC) of Federal University of Minas Gerais (UFMG). He has degree (2003) and Masters (2007) in Computer Science from the Pontifical Catholic University of Minas Gerais (2003). He got his Ph.D. at Federal University of Minas Gerais (UFMG) with a one year period as a visiting researcher in Carnegie Mellon University and a five months period as a visiting researcher at INRIA Lyon. His research interest is mostly focused on knowledge discovery and data mining in complex and distributed systems.
\end{IEEEbiography}

\vspace{-1cm}

\begin{IEEEbiography}[{\includegraphics[width=1in,height=1.25in,clip,keepaspectratio]{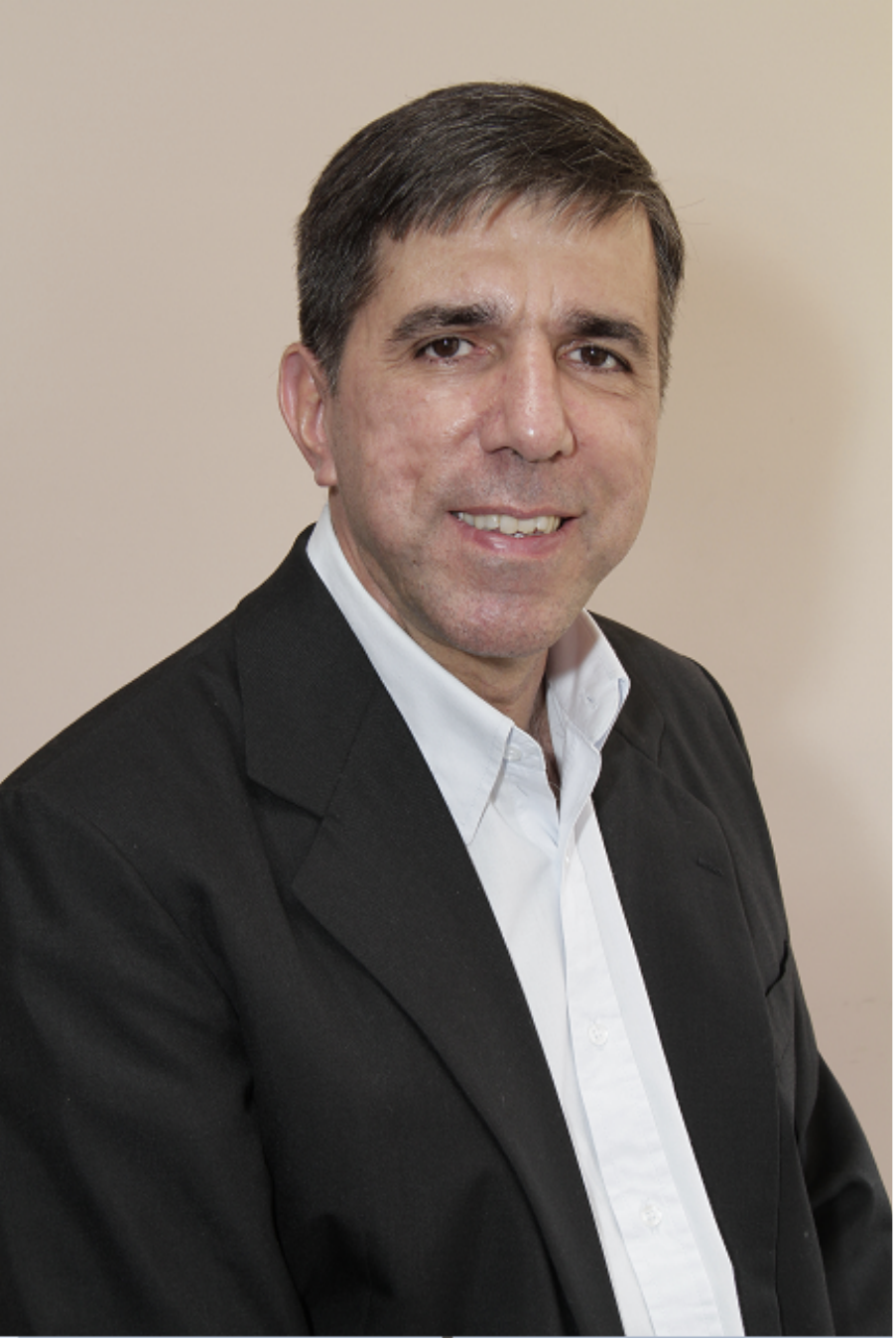}}]{Antonio A. F. Loureiro} received his B.Sc. and M.Sc. degrees in Computer Science from the Federal University of Minas Gerais (UFMG), Brazil, and the Ph.D. degree in Computer Science from the University of British Columbia, Canada. Currently, he is a full professor of Computer Science at UFMG, where he leads the research group in mobile ad hoc networks. His main research areas are mobile computing, vehicular networks, wireless sensor networks, and distributed algorithms. In the last 15 years he has published regularly in international conferences and journals related to those areas, and also presented keynotes and tutorials at international conferences. He was awarded the 2015 IEEE Communications Society Ad Hoc and Sensor Networks Technical Committee recognition award with the citation "for his contributions to the design, modeling and analysis of communication protocols for ad hoc networks".
\end{IEEEbiography}

\vfill

\end{document}